\documentclass[12pt]{article}

\usepackage{euscript,amsmath, amssymb, amsfonts}

\newcommand{\supp}{{\rm supp\,}}
\newcommand{\R}{{\mathbb R}}
\newcommand{\N}{{\mathbb N}}
\newcommand{\V}{{\mathbb V}}
\newcommand{\C}{{\mathbb C}}
\newcommand{\di}{{\rm d}}
\newcommand{\W}[1]{{:\!#1\!\!:}}
\newcommand{\dv}[2]{\langle #1,#2\rangle}

\newcommand{\maj}{w_{\mathrm{maj}}}
\newcommand{\mjz}{\mathbf w_{\mathrm{maj}}}
\newcommand{\wir}{w_{{\scriptscriptstyle IR}}}
\newcommand{\wuv}{w_{{\scriptscriptstyle UV}}}

\newcounter{theorem}
\newcounter{lemma}
\newcounter{defin}
\newcommand{\theorem}{\par\refstepcounter{theorem}{\bf Theorem \arabic{theorem}. }}
\newcommand{\lemma}{\par\refstepcounter{lemma}{\bf Lemma \arabic{lemma}. }}
\newcommand{\defin}{\medskip\par\refstepcounter{defin}{\bf Definition \arabic{defin}. }}

\begin{document}
\pagestyle{myheadings}
\title{Towards Euclidean theory of infrared singular quantum fields}
\author{A.~G.~Smirnov\footnote{E-mail: smirnov@lpi.ru}}
\date{}

\maketitle

\begin{center}
I.~E.~Tamm Theory Department, P.~N.~Lebedev
Physical Institute,\\ Leninsky prospect 53, Moscow 119991, Russia
\end{center}

\bigskip

\begin{center}
Abstract
\end{center}
A new generalized
formulation of the spectral condition is proposed for quantum fields with
highly singular infrared behavior whose vacuum correlation functions are
well defined only under smearing with analytic test functions in
momentum space. The Euclidean formulation of QFT developed by Osterwalder
and Schrader is extended to theories with infrared singular indefinite
metric. The corresponding generalization of the reconstruction theorem is
obtained. The fulfilment of the generalized spectral condition is verified
for quantum fields representable by infinite series in the Wick powers
of indefinite metric free fields.
\thispagestyle{empty}

\newpage
\pagenumbering{arabic}
\section{Introduction}

The Euclidean methods are central to the rigorous
construction of quantum field models with polynomial interaction in
lower dimensions, see~\cite{GJ}.
This construction heavily relies on the use of Osterwalder--Schrader
reconstruction theorem~\cite{OS1,OS2} which allows to
pass from Euclidean Green's functions to quantum field theory in the
Minkowski space-time.
However, the results of~\cite{OS1,OS2} in their initial form
are inapplicable to models with a singular infrared behavior violating
the positivity condition and, in particular, to gauge theories.  The
problem of the Euclidean formulation of QFT in the case of pseudo-Wightman
axioms with an indefinite metric~\cite{MorStr,BLOT} was considered
in~\cite{JakStr} within the traditional framework of tempered Schwartz
distributions. However, as shown, in particular, by the example of the Schwinger model
in an arbitrary $\alpha$-gauge~\cite{CapFer}, the exact operator
solutions of gauge models can be much more singular
and, in general, are
well defined only under smearing with analytic test functions in
momentum space. In this work, we study the possibility of extending the
Euclidean theory to the fields whose vacuum expectation values are analytic
functionals in momentum representation.

One of the main difficulties is connected with the adequate generalization
of the spectral condition, which determines the analyticity properties of
the Wightman functions. In indefinite metric field theories, the
space-time translations are implemented by
pseudo-unitary (in general, unbounded) operators in the state space
and, therefore, the spectral condition
can be formulated only in the weak form, i.e., as a restriction on the
$n$-point Wightman functions $w_n$. When $w_n$ are tempered
distributions, it is of the same form as in the ordinary Wightman
theory:\footnote{To avoid inessential technical complications,
we consider the case of a
single scalar field in $d$-dimensional space-time ($d\geq 2$).}
\begin{equation}
\supp \hat W_n\subset \bar
\V_+^n, \quad \hat W_n(p_1,\ldots,p_n)= \int
W_n(\xi_1,\ldots,\xi_n)e^{ip_1\xi_1+\ldots+ip_n\xi_n}\,\di p_1 \ldots \di
p_n,
\label{i2}
\end{equation}
where $\bar\V_+$ is the closed upper light cone and
$W_n(\xi)$ is the Wightman function $w_{n+1}$ expressed in terms of
the difference variables $\xi_j=x_j-x_{j+1}$:
\begin{equation}
w_{n+1}(x_1,\ldots,x_{n+1})=W_n(x_1-x_2,\ldots,x_n-x_{n+1}).
\label{i3}
\end{equation}
If $\hat W_n$ are analytic functionals, then the
condition~(\ref{i2}) becomes inapplicable because of the lack of test
functions with compact support. The problem of the appropriate
generalization of the spectral condition was raised by Moschella and
Strocchi~\cite{MoschStroc}.
In~\cite{Soloviev3},
it was suggested to overcome this difficulty using the notion of
carrier cone which replaces the notion of support for analytic functionals
and whose existence for a wide class of functionals was proved
in~\cite{Sol1,Sol2}.
The generalized spectral condition which is obtained from~(\ref{i2})
by replacing the support with a carrier cone is sufficient for deriving
the usual analyticity properties of the Wightman
functions~\cite{Soloviev3} and is fulfilled for the sums of infinite
series in the Wick powers of indefinite metric free fields~\cite{SS}.
Moreover, fields representable by such series satisfy even stronger
condition stated in terms of the notion of {\it strong carrier cone} which is introduced
by Definition~\ref{d2} below. The latter arises naturally when one tries to
bring the definition of carrier cone into line with with the operation of
tensor product of functionals, which plays an important role in the problem
under consideration. The use of the generalized spectral condition in such a
stronger form yields simple and effective estimates for the Schwinger
functions which allow developing the Euclidean formulation in complete analogy
with the case of tempered fields~\cite{OS2}.
In this paper, the analysis of the Euclidean formulation of QFT
is
performed  at the level of the Wightman functions of the
theory. At the same time, we believe that the developed approach provides
a basis for considering more subtle questions connected with
finding the Hilbert majorant of an indefinite
metric~\cite{MorStr}.

As in~\cite{MoschStroc}--\cite{Sol2}, we use the Gelfand--Shilov
spaces $S^\alpha_\beta$ as the functional
domain of definition of fields in momentum space.  The generalized
functions belonging to $S^{\prime\alpha}_\beta$
\footnote{Here and subsequently, we
denote the continuous dual of a topological vector space by the same symbol with a
prime.} grow at infinity like
$\exp(|p|^{1/\beta})$ and their Fourier-transforms
like $\exp(|x|^{1/\alpha})$. Thus, the indices $\alpha$ and $\beta$
control, respectively, the possible infrared and ultraviolet singularities.
If $0\leq\alpha<1$, then the elements of $S^\alpha_\beta$ are entire
analytic functions. It is important that our treatment covers the case
$\alpha=0$ which corresponds to an arbitrary singular infrared behavior.

In Sec.~\ref{c1s2},
we introduce the definition of strong carrier cone and prove
that the intersection of strong carrier cones of a functional
is again its strong carrier cone. Analogous result for carrier cones
ensuring, in particular, the existence of the smallest carrier cone was
established in~\cite{Sol1}. In Sec.~\ref{c1s3}, we prove
that the definition of strong carrier cone is compatible with
the operation of tensor product of functionals.
In Sec.~\ref{c1s4}, the theory of Laplace transformation
is extended to functionals with convex strong carrier cones.
In particular, we prove a Paley--Wiener-Schwartz type theorem characterizing those
analytic functions that are Laplace transforms of such
functionals. In Sec.~\ref{c1s5},
this theorem is applied to derive estimates for the Schwinger
functions. In the same section the main result of the paper is presented,
namely, the generalized Euclidean reconstruction theorem which
covers field theories with arbitrarily singular infrared behavior.
In Sec.~\ref{c1s6}, we check that the generalized spectral condition is
satisfied for normally ordered entire functions of indefinite metric free
fields. Sec.~\ref{c1s7} is devoted to concluding remarks.
Some
details of proofs are given in Appendices~A and~B.

\section{Strong carrier cones}
\label{c1s2}

The space $S^\alpha_\beta(\R^k)$ is by definition~\cite{GS}
the union (inductive limit) with respect to $A,B>0$ of the
Banach spaces composed of smooth functions on
$\R^k$ with the finite norm
\begin{equation}
|||f|||_{A,B}=
\sup_{p\in \R^k,\,\lambda,\,\mu}\frac{|p^\mu\partial^\lambda f(p)|}
{A^{|\lambda|} B^{|\mu|} |\lambda|^{\alpha |\lambda|} |\mu|^{\beta|\mu|}},
\label{0}
\end{equation}
where
$\lambda$ and $\mu$ run over all multi-indices and the standard multi-index
notation is used.
The spaces $S^\alpha_\beta$ are nontrivial if $\alpha+\beta>1$ or if
$\alpha>0$ and $\alpha+\beta=1$. From now on, we assume that
one of these conditions is
satisfied. If $0\leq \alpha<1$, then
$S^\alpha_\beta$ consists of (the restrictions to $\R^k$ of)
entire analytic functions and
an alternative description of these spaces
in terms of complex variables is possible~\cite{GS}. Namely, an analytic function
$f$ on $\C^k$
belongs to the class $S^\alpha_\beta$
if and only if
$$
|f(w)|\leq C\exp(-|p/B|^{1/\beta}+|Aq|^{1/(1-\alpha)}),\quad
w=p+iq\in \C^k,
$$
for some $A,B>0$ depending on $f$. For definiteness, we assume the norm
$|\cdot|$ on $\R^k$ to be uniform, i.e.,  $|p|=\sup_{1\leq j\leq
k}|p_j|$. The main element of the approach developed in~\cite{Sol1,Sol2}
is the employment, in addition to the spaces $S^\alpha_\beta$,
of analogous spaces associated with cones.

\defin
\label{d1}
Let $U$ be a nonempty cone in $\R^k$.
The Banach space $S^{\alpha,A}_{\beta,B}(U)$, $0\leq \alpha<1$,
consists of entire analytic functions on
$\C^k$ with the finite norm
\begin{equation}
\|f\|_{U,A,B}= \sup_{w\in \C^k} |f(w)|
\exp(|p/B|^{1/\beta}-|Aq|^{1/(1-\alpha)}-
\delta_U(Ap)^{1/(1-\alpha)}),
\label{1}
\end{equation}
where
$\delta_U(p)=\inf_{p'\in U}|p-p'|$.
The space $S^{\alpha}_{\beta}(U)$ is defined to be the inductive limit
$\varinjlim_{A,B>0}S^{\alpha,A}_{\beta,B}(U)$.

A nonempty closed cone $K$ is called a carrier cone of the functional
$u\in S^{\prime \alpha}_{\beta}(\R^k)$ if $u$ extends continuously
to every space
$S^{\alpha}_{\beta}(U)$, where $U$ is a cone with an open
projection\footnote{The projection of the cone $U$
is by definition the intersection of $U$ with the unit
sphere in $\R^k$; the projection of $U$ is meant
to be open in the topology of the sphere.} such that $K\subset U$.
As shown in~\cite{Sol1,Sol2}, the space $S^{\alpha}_{\beta}(\R^k)$ is
dense in each space $S^{\alpha}_{\beta}(U)$.  The space of the
functionals carried by the cone $K$ is therefore identified with
$s^{\prime \alpha}_{\beta}(K)$, where
$s^{\alpha}_{\beta}(K)=\varinjlim_{U}S^{\alpha}_{\beta}(U)$.

It should be mentioned that in~\cite{Sol1,Sol2}, the spaces
$S^{\alpha}_{\beta}(U)$ are defined for open cones $U$ and a closed cone $K$ is said
to be a carrier cone of $u$ if this functional has a continuous extension to every
$S^{\alpha}_{\beta}(U)$, where $K\setminus\{0\}\subset U$.
This definition is equivalent to the one given here.
It is easy to see that all results of~\cite{Sol1,Sol2} concerning the spaces
$S^{\alpha}_{\beta}(U)$
remain true for any nonempty cone $U$.
In what follows, we find it
convenient to use the spaces $S^{\alpha}_{\beta}(U)$ associated with arbitrary nonempty
cones because this
allows handling the
degenerate cone $\{0\}$ on the same footing as nondegenerate closed carrier cones. We
also note that in~\cite{Sol1,Sol2}, the space $s^{\alpha}_{\beta}(K)$
was denoted by $S^{\alpha}_{\beta}(K)$. Here, such notation might lead
to confusion because the spaces $S^{\alpha}_{\beta}(K)$ and
$S^{\alpha}_{\beta}(U)$ are no longer distinguished by the type of the
cone.

The following result established in~\cite{Sol1} shows that every
functional of the class $S^{\prime \alpha}_{\beta}(\R^k)$ has a uniquely
defined minimal carrier cone.

\theorem
\label{t1}
{\it If both $K_1$ and $K_2$ are carrier cones of
$u\in S^{\prime \alpha}_{\beta}(\R^n)$, then so is $K_1\cap K_2$.}

The cone $\bar\V^n_+$, which enters into the formulation~(\ref{i2}) of
the spectral condition, has a natural direct product structure and
the following definition turns out to be useful for generalizing the
spectral condition.
\defin
\label{d2}
Let $K_1,\ldots,K_n$ be nonempty closed cones in
$\R^{k_1},\ldots,\R^{k_n}$ respectively. The cone
$K_1\times\ldots\times K_n$
is called a strong carrier cone of the functional $u\in S^{\prime
\alpha}_{\beta}(\R^{k_1+\ldots+k_n})$ if $u$ allows a continuous
extension to the space
$s^{\alpha}_{\beta}(K_1,\ldots,K_n)=
\varinjlim_{U_1,\ldots,U_n}S^{\alpha}_{\beta}(U_1\times\ldots\times U_n)$,
where the inductive limit is taken over all cones
$U_1,\ldots,U_n$ with open projections such that $K_j\subset U_j$ for all $j=1,\ldots,n$.

The meaning of the spaces $s^{\prime\alpha}_{\beta}(K_1,\ldots,K_n)$
is clarified by Lemma~\ref{l5} below.
If $n=1$, then we recover the definition of carrier cone.
As shown in~\cite{Sol2}, the natural embeddings
$S^{\alpha,A}_{\beta,B}(U)\to S^{\alpha,A'}_{\beta,B'}(U)$
are compact for $A'>A,\, B'>B$ sufficiently large.
Therefore, $S^\alpha_\beta(U)$ and
$s^{\alpha}_{\beta}(K_1,\ldots,K_n)$
are DFS-spaces (we recall that DFS-spaces are, by definition,
the inductive limits of injective compact sequences of
locally convex spaces).
In particular, they (and their duals) are reflexive, complete, and
Montel spaces~\cite{Komatsu}.

Clearly, $s^{\prime\alpha}_{\beta}(K_1,\ldots,K_n)\subset
s^{\prime\alpha}_{\beta}(K_1\times\ldots\times K_n)$, but
the following example shows that the condition
$u\in s^{\prime\alpha}_{\beta}(K_1,\ldots,K_n)$ is, in general, stronger
than the condition
$u\in s^{\prime\alpha}_{\beta}(K_1\times\ldots\times K_n)$.

{\bf Example 1.} Let $u(p)$ be the function equal to unity on the set
$\{p\in \R^2\,|\,p_2\geq -|p_1|^{2/3}\}$ and zero outside
this set.
As a generalized function,
$u$ obviously belongs to
$s^{\prime\alpha}_{\beta}(\R\times\bar\R_+)$ for all $0\leq\alpha<1$ and
$\beta>1-\alpha$. Let us show that $u\notin
s^{\prime 2/3}_{1/2}(\R,\bar\R_+)=S^{\prime 2/3}_{1/2}(\R\times\R_+)$.
Use the test function $f(w)=\exp(-w_1^2-w_2^3)$ belonging to
$S^{2/3}_{1/2}(\R\times\R_+)$ and define $g(w)$ by the same formula as
$f$ but with the twice less exponent.
By the above-mentioned density property, there
exists a sequence
$g_{\nu}\in S^{2/3}_{1/2}(\R^2)$ converging to $g$ in
$S^{2/3}_{1/2}(\R\times\R_+)$. Set $f_{\nu}(w)=g_{\nu}(w)\overline
{g_{\nu}(\bar w)}$ (bar means complex conjugation). Obviously,
$f_{\nu}(p)\geq 0$ and $f_{\nu}\to f$ in $S^{2/3}_{1/2}(\R\times\R_+)$.
Since the integral $\int u(p) f(p)\di p$
is divergent, we conclude by the monotonic convergence theorem that
$\int u(p) f_{\nu}(p)\di p\to\infty$ as $\nu\to\infty$.

The following analogue of Theorem~\ref{t1} is valid.

\theorem
\label{t2}
{\it Let
$K_1^{(1,2)},\ldots,K_n^{(1,2)}$ be nonempty closed cones in
$\R^{k_1},\ldots,\R^{k_n}$ respectively. If both
$K_1^{(1)}\times\ldots\times K_n^{(1)}$ and
$K_1^{(2)}\times\ldots\times K_n^{(2)}$
are strong carrier cones of $u\in S^{\prime
\alpha}_{\beta}(\R^{k_1+\ldots+k_n})$, then so is $(K_1^{(1)}\cap
K_1^{(2)})\times\ldots\times (K_n^{(1)}\cap K_n^{(2)})$.}

Before we pass to the proof, let us set up the
notation and recall some facts concerning cones in $\R^k$.  Let ${\mathcal
C}(\R^k)$ denote the set of all cones in $\R^k$ containing the origin and
let ${\mathcal O}(\R^k)$ be the subset of ${\mathcal C}(\R^k)$ consisting
of cones with open projections (or, which is the same, of those cones whose
intersection with $\R^k\setminus\{0\}$ is open).  We note that the cones
$U_j$ in Definition~\ref{d2} belong to
${\mathcal O}(\R^{k_j})$.  Obviously, for any (open) subset $O$ of the unit
sphere there is a unique cone $U\in {\mathcal C}(\R^k)$ (resp.,
$U\in {\mathcal O}(\R^k)$) such that $O$ is the projection of $U$.  Using
this one-to-one correspondence, one can apply standard
compactness arguments to cones in ${\mathcal C}(\R^k)$ to obtain:
\begin{itemize}
\item[(I)] if $U\in {\mathcal C}(\R^k)$, $V\in {\mathcal O}(\R^k)$, and
$U\Subset V$,\footnote{A cone $U$ is said to be compact in a cone $V$ (notation $U\Subset V$)
if $\bar U\setminus\{0\}\subset V$.} then
there exists $W\in{\mathcal O}(\R^k)$ such that $U\Subset W\Subset V$;
\item[(II)] if $U_1,U_2\in {\mathcal C}(\R^k)$ and $\bar U_1\cap\bar
U_2=\{0\}$, then there exist $V_1,V_2\in {\mathcal O}(\R^k)$ such that
$U_{1,2}\Subset V_{1,2}$ and $\bar V_1\cap \bar V_2=\{0\}$;
\item[(III)] if
$V\in {\mathcal C}(\R^k)$, $U\in {\mathcal O}(\R^k)$, and $V\Subset U$,
then $\bar V\cap \Delta U=\{0\}$, where $\Delta U= (\R^k\setminus
U)\cup\{0\}$ corresponds to the complement of the projection of $U$ in
the unit sphere.
\end{itemize}

{\it Proof of Theorem~$\ref{t2}$}.
Let $u_{1,2}$ be the extensions of $u$ to the spaces
$s^\alpha_\beta(K_1^{(1,2)},\ldots,K_n^{(1,2)})$ which exist by the
hypothesis and let
$f\in s^\alpha_\beta(K_1^{(1)},\ldots,K_n^{(1)})\cap
s^\alpha_\beta(K_1^{(2)},\ldots,K_n^{(2)})$. By
Definition~\ref{d2}, there are cones
$U_1^{(1,2)}\in {\mathcal O}(\R^{k_1}),\ldots,U_n^{(1,2)}\in
{\mathcal O}(\R^{k_n})$ such that
$K_1^{(1,2)}\times\ldots\times K_n^{(1,2)}\subset
V^{(1,2)}= U_1^{(1,2)}\times\ldots\times U_n^{(1,2)}$ and $f\in
S^\alpha_\beta(V^{(1)})\cap
S^\alpha_\beta(V^{(2)})=S^\alpha_\beta(V^{(1)}\cup V^{(2)})$. The existence
of continuous dense embeddings
$S^\alpha_\beta(\R^{k_1+\ldots+k_n})\to
S^\alpha_\beta(V^{(1)}\cup V^{(2)})\to
s^\alpha_\beta(K_1^{(1,2)},\ldots,K_n^{(1,2)})$ implies that $u_1$
and $u_2$ coincide on $S^\alpha_\beta(V^{(1)}\cup V^{(2)})$ and,
consequently,
\begin{equation} u_1(f)=u_2(f) \mbox{ for every } f\in
s^\alpha_\beta(K_1^{(1)},\ldots,K_n^{(1)})\cap
s^\alpha_\beta(K_1^{(2)},\ldots,K_n^{(2)}).
\label{lab1a}
\end{equation}
Let us consider the mapping
$$
j:\, s^{\alpha}_{\beta}(K_1^{(1)},\ldots,K_n^{(1)})\times
s^{\alpha}_{\beta}(K_1^{(2)},\ldots,K_n^{(2)}) \to
s^{\alpha}_{\beta}(K_1^{(1)}\cap K_1^{(2)},\ldots,K_n^{(1)}\cap K_n^{(2)})
$$
taking $(f_1,f_2)$ to $f_1-f_2$. If $j$ is surjective, then
$s^{\alpha}_{\beta}(K_1^{(1)}\cap K_1^{(2)},\ldots,K_n^{(1)}\cap K_n^{(2)})$
is topologically isomorphic to
the quotient space $[s^{\alpha}_{\beta}(K_1^{(1)},\ldots,K_n^{(1)})\times
s^{\alpha}_{\beta}(K_1^{(2)},\ldots,K_n^{(2)})]/\ker j$
by the open mapping theorem (see~\cite{Schaefer}, Theorem IV.8.3.), which is
applicable because all spaces under consideration are DFS.
From (\ref{lab1a}), it follows that $\ker j$ is contained in the kernel of
the functional $(f_1,f_2)\to u_1(f_1)+u_2(f_2)$. As a consequence, the
latter allows a canonical decomposition of the form $\tilde u\circ j$,
where $\tilde u$ belongs to $s^{\prime\alpha}_{\beta}(K_1^{(1)}\cap
K_1^{(2)},\ldots,K_n^{(1)}\cap K_n^{(2)})$ and, as one can easily see, is
the extension of $u$. Thus, it remains to prove the
surjectivity of $j$.
It is ensured by the following
decomposition theorem for test functions.

\theorem
\label{t3}
{\it If
$f\in s^{\alpha}_{\beta}(K^{(1)}_1\cap
K^{(2)}_1,\ldots,K^{(1)}_n\cap K^{(2)}_n)$ then
$f=f^{(1)}+f^{(2)}$ with $f^{(1,2)}\in
s^{\alpha}_{\beta}(K^{(1,2)}_1,\ldots,K^{(1,2)}_n)$.}

To prove Theorem~\ref{t3}, we need three lemmas.

\lemma
\label{l1}
{\it Let $U\in {\mathcal C}(\R^{k_1})$, $V\in {\mathcal C}(\R^{k_2})$, and
let $U_1,U_2\in {\mathcal C}(\R^{k_1})$ be such that $\bar U_1\cap\bar
U_2=\{0\}$. Then for every $f\in S^{\alpha}_{\beta}(U\times V)$
one can find $f_{1,2}\in S^{\alpha}_{\beta}((U\cup
U_{1,2})\times V)$ such that
$f=f_1+f_2$. If $U\in {\mathcal O}(\R^{k_1})$, then
the condition $\bar U_1\cap\bar U_2=\{0\}$ can be replaced by
$\bar U_1\cap\bar U_2\subset U$.

Proof} will be given for $0<\alpha<1$, when the space
$S^{\alpha}_{1-\alpha}$ is nontrivial. The more difficult case $\alpha=0$
is considered in Appendix~A.
By (II), there exist cones
$Q_1,Q_2\in {\mathcal O}(\R^{k_1})$ such that
$U_{1,2}\Subset Q_{1,2}$ and
$\bar Q_1\cap\bar
Q_2=\{0\}$, and in view of (I)
one can find cones $V_1,V_2\in {\mathcal
O}(\R^{k_1})$ such that $U_{1,2}\Subset V_{1,2}\Subset Q_{1,2}$.
Set $W_1=Q_1$ and
$W_2=\Delta Q_1$. By (III) we have $\bar W_1\cap \bar V_2=\bar
V_1\cap \bar W_2=\{0\}$. Let $g_0\in
S^{\alpha,A_0}_{1-\alpha,B_0}(\R^{k_1})$ and $\int_{\R^{k_1}}g_0(p')\di
p'=1$.  We set
\begin{equation}
g_{1,2}(w)=\int_{W_{2,1}}
g_0(w'-\eta)\di\eta,\quad w=(w',w'')\in \C^{k_1}\times \C^{k_2}.
\label{1b}
\end{equation}
Obviously, $g_1,g_2$ are entire analytic functions on
$\C^{k_1}\times \C^{k_2}$ and $g_1+g_2=1$.
If $\eta\in W_{1,2}$, then $|p'-\eta|\geq \delta_{W_{1,2}}(p')$ and in view
of~(\ref{1}) we have
\begin{equation}
|g_{1,2}(w)|\leq C\exp\left
[(A_0|q'|)^{1/(1-\alpha)}-\delta_{W_{2,1}}
\left(\frac{p'}{2B_0}\right)^{1/(1-\alpha)}\right].
\label{2}
\end{equation}
Set $f_{1,2}=fg_{1,2}$ and fix
$A,B>0$ such that
$f\in S^{\alpha,A}_{\beta,B}(U\times V)$.
If $p'\notin V_1$, then
$\delta_{U_1}(p')\geq \theta|p'|$ for some $0<\theta<1$ and in view of
the inequality $|p'|\geq \delta_U(p')$ we have
$\delta_U(p')\leq\delta_{U_1\cup U}(p'/\theta)$. Hence,
taking
(\ref{1}), (\ref{2}) and the relation $\delta_{U\times
V}(p)=\max[\delta_U(p'),\delta_V(p'')]$
into account, we find that
\begin{equation}
|f_1(w)| \leq C\|f\|_{A,B} \exp\left
[(2(A_0+A)|q|)^{\frac{1}{1-\alpha}}+
\delta_{(U_1\cup U)\times V}\left(Ap/\theta\right)^{\frac{1}{1-\alpha}}
-\left(|p|/B\right)^{1/\beta}\right]
\label{3}
\end{equation}
for $p'\notin V_1$.
Further, there is $\theta_1>0$ such that
$\delta_{W_2}(p')\geq \theta_1 |p'|$ for $p'\in V_1$. Therefore, for
$A\leq\theta_1/2B_0$, using (\ref{1}) and (\ref{2}), we obtain
\begin{equation}
|f_1(w)|
\leq C \|f\|_{A,B}\exp\left
[(2(A_0+A)|q|)^{\frac{1}{1-\alpha}}+
\delta_V(Ap'')^{\frac{1}{1-\alpha}}
-(|p|/B)^{1/\beta}\right]
\label{4}
\end{equation}
for $p'\in V_1$. Comparing (\ref{3}) and (\ref{4}), we conclude that
$f_{1}\in S^{\alpha,A'}_{\beta,B}((U\cup U_1)\times V)$ for
$A'\geq 2(A_0+A)+A/\theta$. Analogous arguments show that
$f_{2}\in S^{\alpha}_{\beta}((U\cup U_2)\times V)$ for $B_0$ sufficiently
large.

If $U\in {\mathcal O}(\R^{k_1})$, then $K_1\cap K_2=\{0\}$ for the nonempty
closed cones
$K_{1,2}=\bar
U_{1,2}\cap \Delta U$, and according to the above we have a decomposition
$f=f_1+f_2$, where $f_{1,2}\in S^{\alpha}_{\beta}((U\cup
K_{1,2})\times V)$. To complete the proof, it remains to note that
$K_{1,2}\cup U\supset U_{1,2}\cup U$.

\lemma
\label{l2}
{\it Let
$U_1\in {\mathcal O}(\R^{k_1})$, $U_2\in {\mathcal O}(\R^{k_2})$,
$U\in {\mathcal C}(\R^k)$,
and $V_{1,2}\in {\mathcal C}(\R^{k_{1,2}})$ be such that
$V_{1,2}\Subset U_{1,2}$.  Then for every $f\in
S^{\alpha}_{\beta}(U_1\times U_2\times U)$, there is a decomposition
$f=f_1+f_2$, where $f_{1}\in S^{\alpha}_{\beta}(V_1\times \R^{k_2}\times
U)$ and $f_{2}\in S^{\alpha}_{\beta}(\R^{k_1}\times V_2\times U)$.

Proof.}  By (I), one can find
$W_{1,2}\in{\mathcal O}(\R^{k_{1,2}})$ such that
$V_{1,2}\Subset W_{1,2}\Subset U_{1,2}$.  Set $Q_1=\bar V_1\times
\Delta W_2$ and $Q_2=\Delta W_1\times \bar V_2$.
According to (III) we have $\bar Q_1\cap \bar Q_2=\{0\}$
and by Lemma~\ref{l1},
$f=f_1+f_2$, where $f_{1,2}\in S^{\alpha}_{\beta} ([(U_1\times
U_2)\cup Q_{1,2}]\times U)$. It remains to note that
$(U_1\times U_2)\cup Q_1\supset V_1\times \R^{k_2}$ and $(U_1\times U_2)\cup
Q_2\supset \R^{k_1}\times V_2$.

\lemma
\label{l3}
{\it Let $V_1,U_1,\ldots,V_n,U_n$ be cones in
$\R^{k_1},\ldots, \R^{k_n}$ such that
$V_j\in {\mathcal C}(\R^{k_j})$,
$U_j\in {\mathcal O}(\R^{k_j})$, and $V_j\Subset U_j$ for all
$j=1,\ldots,n$. If $f\in
S^{\alpha}_{\beta}(U_1\times\ldots\times U_n)$, then
$f=f_1+\ldots+f_n$ with $f_j\in
S^{\alpha}_{\beta}(\R^{k_1}\times\ldots\times \R^{k_{j-1}}\times
V_j\times \R^{k_{j+1}}\times\ldots\times \R^{k_n})$.

Proof.} We shall prove the following stronger statement.
Let the cones $V_1,U_1,\ldots,V_n,U_n$ satisfy the conditions of the lemma
and let
$U\in{\mathcal C}(\R^k)$. Then for every $f\in
S^{\alpha}_{\beta}(U_1\times\ldots\times U_n\times U)$ there exists a
decomposition
$f=f_1+\ldots+f_n$, where $f_j\in
S^{\alpha}_{\beta}(\R^{k_1}\times\ldots\times \R^{k_{j-1}}\times
V_j\times \R^{k_{j+1}}\times\ldots\times \R^{k_n}\times U)$.
The statement of the lemma corresponds to the particular case
$\R^k=U=\{0\}$. For $n=2$, the proof is reduced to applying Lemma~\ref{l2}.
For $n>2$, we proceed by induction on $n$. Supposing the statement to
hold up to $n-1$, we choose
the cones $W_{1,2}\in {\mathcal
O}(\R^{k_{1,2}})$ such that $V_{1,2}\Subset W_{1,2}\Subset U_{1,2}$.
By Lemma~\ref{l2}, $f=\tilde f_1+\tilde
f_2$, where $\tilde f_1 \in S^{\alpha}_{\beta}(W_1\times \R^{k_2}\times
U_3\times\ldots\times U_n\times U)$ and $\tilde f_2 \in
S^{\alpha}_{\beta}(\R^{k_1}\times W_2\times U_3\times\ldots\times U_n\times
U)$, and in view of the natural isomorphisms
$W_1\times \R^{k_2}\times U_3\times\ldots\times U_n\times U\cong W_1\times
U_3\times\ldots\times U_n\times (\R^{k_2}\times U)$ and $\R^{k_1}\times
W_2\times U_3\times\ldots\times U_n\times U\cong W_2\times
U_3\times\ldots\times U_n\times (\R^{k_1}\times U)$ we obtain the
desired decompositions of $\tilde f_1$ and $\tilde f_2$. The lemma is
proved.

We now prove Theorem~\ref{t3}.
By Lemma~\ref{l3}, we have a decomposition
$f=f_1+\ldots+f_n$, where $f_j\in
s^{\alpha}_{\beta}(\R^{k_1},\ldots,\R^{k_{j-1}},K^{(1)}_j\cap K^{(2)}_j,
\R^{k_{j+1}},\ldots,\R^{k_n})$, $j=1,\ldots,n$. Let the cones
$U^{(1)}_j, U^{(2)}_j, U_j\in {\mathcal O}(\R^{k_j})$ be such that
$f_j\in S^{\alpha}_{\beta}(\R^{k_1},\ldots,\R^{k_{j-1}},U_j,
\R^{k_{j+1}},\ldots,\R^{k_n})$, $K^{(1,2)}_j\Subset U^{(1,2)}_j$ and $\bar
U^{(1)}_j\cap \bar U^{(2)}_j\subset U_j$.  By Lemma~\ref{l1}
there is a decomposition $f_j=f_j^{(1)}+f_j^{(2)}$, where $f_j^{(1,2)}\in
S^{\alpha}_{\beta}(\R^{k_1}\times\ldots\times \R^{k_{j-1}}\times
U_j^{(1,2)}\times \R^{k_{j+1}}\times\ldots\times \R^{k_n})$. Setting
$f^{(1,2)}=f_1^{(1,2)}+\ldots+f_n^{(1,2)}$, we arrive at the desired
result.

\section{Tensor products}
\label{c1s3}

We refer the reader to~\cite{Schaefer} for the definition and properties of
the inductive topology ($i$-topology), projective topology
($\pi$-topology), and the topology of equicontinuous convergence
($e$-topology) on tensor products of locally convex spaces.
Recall
that $\pi$- and $e$-topologies coincide on the tensor products of
nuclear spaces, while $i$- and $\pi$-topologies coincide on the tensor products
of Fr\'echet spaces.

\lemma
\label{l4}
{\it Let $L^{(1)}$ and $L^{(2)}$ be {\rm DFS}-spaces. Then
$L^{(1)}\otimes_i L^{(2)}=L^{(1)}\otimes_\pi L^{(2)}$.
If $L^{(1)}$ is nuclear, then
$(L^{(1)}\hat\otimes_i L^{(2)})'=
L^{(1)\prime}\hat\otimes_i
L^{(2)\prime}$,
where the hat means completion and the prime denotes the strong dual space.}

The proof is given in Appendix~B. In~\cite{Treves}, it was shown that
if $L^{(1)}$, $L^{(2)}$, and $L$ are the strong duals of reflexive
Fr\'echet spaces, then every separately continuous bilinear map of
$L^{(1)}\times L^{(2)}$ into $L$ is continuous. From
Lemma~\ref{l4}, it follows that if $L^{(1)}$ and $L^{(2)}$ are DFS-spaces,
then an analogous statement holds for any locally convex space~$L$.

Let $L^{(1)},\ldots, L^{(n)}$ be locally convex spaces.
We denote by $L^{(1)}\hat\otimes_i\ldots\hat\otimes_i L^{(n)}$
the completion of $L^{(1)}\otimes\ldots\otimes L^{(n)}$
relative to $i$-topology.
If $L_n$ is a barrelled space, then there is the canonical identification
\begin{equation}
L^{(1)}\hat\otimes_i\ldots\hat\otimes_i
L^{(n)} = (L^{(1)}\hat\otimes_i\ldots\hat\otimes_i L^{(n-1)})\hat\otimes_i
L^{(n)}
\label{6}
\end{equation}
(to construct this isomorphism, one can use theorems III.5.2 and
III.5.4 of~\cite{Schaefer}).

\lemma
\label{l5}
{\it Let $K_1,\ldots, K_n$ be nonempty closed cones in
$\R^{k_1},\ldots \R^{k_n}$ respectively and let $0\leq \alpha<1$.
Then we have the isomorphisms
\begin{align} &
s^{\alpha}_\beta(K_1,\ldots,K_n)=s^{\alpha}_\beta(K_1,\ldots,K_{n-1})
\hat\otimes_i s^{\alpha}_\beta(K_n),\nonumber\\
&s^{\prime\alpha}_\beta(K_1,\ldots,K_n)=
s^{\prime\alpha}_\beta(K_1)\hat\otimes_i\ldots\hat\otimes_i
s^{\prime\alpha}_\beta(K_n).\nonumber
\end{align}

Proof.}
As we have already mentioned above, the spaces introduced by
Definition~\ref{d2} are DFS. Moreover, they are nuclear as
countable inductive limits of the spaces $S^{\alpha}_{\beta}(U)$,
whose nuclearity was established in~\cite{Sol2}.
Since $s^{\alpha}_\beta(K_1,\ldots,K_n)$ is Hausdorff and
complete~\cite{Komatsu}, the first isomorphism follows
immediately from Definition~\ref{d2} and the existence of the natural
identification
$S^{\alpha}_\beta(U_1\times U_2)= S^{\alpha}_\beta(U_1)\hat\otimes_i
S^{\alpha}_\beta(U_2)$ for any nonempty cones $U_1, U_2$, see~\cite{Sol2}, Theorem~3.
The second isomorphism
is obtained by induction from the first one using
(\ref{6}) and Lemma~\ref{l4}.

\lemma
\label{l5a}
{\it Let $G_1$ and $G_2$ be subspaces of locally convex spaces
$L_1$ and $L_2$ respectively. Then the topology of equicontinuous
convergence on
$G_1\otimes G_2$ coincides with that induced from
$L_1\otimes_e L_2$.

Proof.} Let $j_{1,2}$ be the natural injections of $G_{1,2}$ into
$L_{1,2}$ and let $j=j_1\otimes j_2$. We denote by ${\mathcal E}_{1,2}$
($\tilde{\mathcal E}_{1,2}$) the families of equicontinuous subsets of
$L'_{1,2}$ (resp., of $G'_{1,2}$). The polar sets of $(S_1\otimes
S_2)^\circ$, $S_{1,2}\in {\mathcal E}_{1,2}$ form the basis of
neighborhoods of the origin for $e$-topology on $L_1\otimes L_2$. Since
$j'_{1,2}({\mathcal E}_{1,2})=\tilde{\mathcal E}_{1,2}$ according
to~\cite{Schaefer}, Theorem IV.4.1, the sets $[j'_1(S_1)\otimes j'_2(S_2)]^\circ$,
$S_{1,2}\in {\mathcal E}_{1,2}$, form the basis of neighborhoods of the origin
for $e$-topology on $G_1\otimes G_2$. It remains to note that in view
of Proposition IV.2.3a of \cite{Schaefer} and the equality $j'=j'_1\otimes j'_2$ these
sets coincide with $j^{-1}[(S_1\otimes S_2)^\circ]$.

\section{Laplace transformation}
\label{c1s4}

\defin
\label{d3}
Let $\beta>1$, let $V_1,\ldots,V_n$ be nonempty open connected cones in
$\R^{k_1},\ldots,\R^{k_n}$ respectively, and let $V=V_1\times\ldots\times
V_n$. The space ${\mathcal A}^\beta_\alpha(V_1,\ldots,V_n)$ with
$0<\alpha<1$ (with $\alpha=0$) consists of analytic functions
in
$T^V=\R^{k_1}\times\ldots\times \R^{k_n}+iV$ that have,
for any $\epsilon>0$ (resp.,
for any $R,\epsilon>0$), the finite norms
\begin{align}
&\|\mathbf v\|_{\epsilon,V'_1,\ldots,V'_n}=\sup_{z\in T^{V'}} |\mathbf v(z)|
\prod\nolimits_{j=1}^n
\exp(-\epsilon|z_j|^{1/\alpha}-\epsilon|y_j|^{-1/(\beta-1)})
\nonumber \\
& \left(\mbox{resp., }
\|\mathbf v\|_{\epsilon,R,V'_1,\ldots,V'_n}=\sup_{z\in T^{V'},\,|z_j|\leq R}
|\mathbf v(z)| \prod\nolimits_{j=1}^n\exp(-\epsilon|y_j|^{-1/(\beta-1)})
\right), \quad
y_j= \mathrm{Im}\, z_j,
\nonumber
\end{align}
where $V'_1,\ldots,V'_n$ are arbitrary cones compact in $V_1,\ldots,
V_n$ and $V'=V'_1\times\ldots\times V'_n$.

If a nondegenerate bilinear form $\dv{\cdot}{\cdot}$ is fixed on
$\R^k$, then the Fourier transform of a test function
$f(x)\in S^\beta_\alpha(\R^k)$ is defined by $\hat f(p)=\int
f(x) e^{i\dv{p}{x}}\,\di x$.  The mapping $f\to \hat f$ is a topological
isomorphism of $S^\beta_\alpha(\R^k)$ onto $S^\alpha_\beta(\R^k)$.  If
$\beta>1$, $V$ is an open connected cone in $\R^k$, and
$K=V^*=\{p\,:\,\dv{p}{y}\geq 0\,\, \forall y\in V\}$, then, as one can
easily see, $e^{i\dv{\cdot}{z}}\in s^{\alpha}_\beta(K)$  for
any $z\in T^V$. The Laplace transform ${\mathcal L}_V u$ of a functional
$u\in s^{\prime\alpha}_\beta(K)$ is defined by $({\mathcal
L}_V u)(z)=u(e^{i\dv{\cdot}{z}})$, $z\in T^V$. As shown
in~\cite{Sol2}, the Laplace operator
${\mathcal L}_V$ is a topological isomorphism of
$s^{\prime\alpha}_\beta(K)$ onto ${\mathcal
A}^\beta_\alpha(V)$ and hence ${\mathcal A}^\beta_\alpha(V)$
is a reflexive Fr\'echet space.

For
$\R^k=\R^{k_1}\times\ldots\times\R^{k_n}$, we assume that
$\dv{p}{x}=\sum_{j=0}^n \dv{p_j}{x_j}_j$,
where $\dv{\cdot}{\cdot}_j$ is a nondegenerate bilinear form on~$\R^{k_j}$.

\lemma
\label{l6}
{\it Let $\beta>1$, $0\leq \alpha<1$,
and ${\bf
v}\in {\mathcal A}^\beta_\alpha(V_1,\ldots,V_n)$, where
$V_1,\ldots,V_n$ are nonempty open connected cones in
$\R^{k_1},\ldots,\R^{k_n}$ respectively. Then ${\bf
v}(z_1,\ldots,z_{n-1},\cdot)\in {\mathcal A}^\beta_\alpha(V_n)$ for any
$z_1\in T^{V_1},\ldots,z_{n-1}\in T^{V_{n-1}}$ and
${\bf v}_u(z_1,\ldots,z_{n-1})=u({\bf
v}(z_1,\ldots,z_{n-1},\cdot))$ belongs to ${\mathcal
A}^\beta_\alpha(V_1,\ldots,V_{n-1})$ for all $u\in {\mathcal
A}^{\prime\beta}_\alpha(V_n)$. The mapping $u\to {\bf v}_u$ from ${\mathcal
A}^{\prime\beta}_\alpha(V_n)$ into ${\mathcal
A}^\beta_\alpha(V_1,\ldots,V_{n-1})$ is continuous.

Proof.} We define the space
$\mathrm A^\beta_\alpha(V_1,\ldots,V_n)$ in the same way as
${\mathcal A}^\beta_\alpha(V_1,\ldots,V_n)$ with the only
difference that the elements of
$\mathrm A^\beta_\alpha$ need not be analytic
functions. ${\mathcal A}^\beta_\alpha(V_1,\ldots,V_n)$ is a closed
subspace of $\mathrm
A^\beta_\alpha(V_1,\ldots,V_n)$. Let $0<\alpha<1$, let $\epsilon>0$, and
let $V'_1,\ldots,V'_n$ be closed subcones of $V_1,\ldots,V_n$. Set
$B_{\epsilon,V'_n}=\{u\in {\mathcal A}^{\prime\beta}_\alpha(V_n)\,:\,
|u({\bf w})|\leq \|{\bf w}\|_{\epsilon,V'_n}\,\forall {\bf w}\in {\mathcal
A}^{\beta}_\alpha(V_n)\}$.
Using Definition~\ref{d3}, we obtain
\begin{multline}
|u({\bf
v}(z_1,\ldots,z_{n-1},\cdot))|\leq \|{\bf
v}(z_1,\ldots,z_{n-1},\cdot)\|_{\epsilon,V'_n} \leq \\ \leq \|\mathbf
v\|_{\epsilon,V'_1,\ldots,V'_n} \prod\nolimits_{j=1}^{n-1}
\exp(\epsilon|z_j|^{1/\alpha}+\epsilon|y_j|^{-1/(\beta-1)})
\nonumber
\end{multline}
for every $u\in B_{\epsilon,V'_n}$ and every
$z_j\in T^{V'_j}$, $1\leq j\leq n-1$.
Consequently, $\|\mathbf v_u\|_{\epsilon,V'_1,\ldots,V'_{n-1}}\leq
\|\mathbf v\|_{\epsilon,V'_1,\ldots,V'_{n}}$
for $u\in B_{\epsilon,V'_n}$.
Thus, $\mathbf v_u$ belongs to the space
$\mathrm A^\beta_\alpha(V_1,\ldots,V_{n-1})$
for any
$u\in {\mathcal A}^{\prime\beta}_\alpha(V_n)$
and the image of $B_{\epsilon,V'_n}$ under the mapping
$u\to {\bf v}_u$ is bounded in this space. The scalar multiples of
$B_{\epsilon,V'_n}$ form a fundamental system of
bounded subsets in the space ${\mathcal A}^{\prime\beta}_\alpha(V_n)$,
which is bornologic as the strong dual of a Fr\'echet space,
see~\cite{Schaefer}, Sec. IV.6.6. Consequently, the mapping
$u\to{\bf v}_u$ from
 ${\mathcal A}^{\prime\beta}_\alpha(V_n)$ to
$\mathrm A^\beta_\alpha(V_1,\ldots,V_{n-1})$ is continuous.
Let $\delta_z$, $z\in T^{V_n}$, be the functional in ${\mathcal
A}^{\prime\beta}_\alpha(V_n)$ which is equal to $\mathbf w(z)$ on
the test function $\mathbf w\in {\mathcal A}^{\beta}_\alpha(V_n)$.
Since ${\mathcal A}^{\beta}_\alpha(V_n)$ is a reflexive space,
the linear span $L$
of such functionals is dense in ${\mathcal
A}^{\prime\beta}_\alpha(V_n)$.  It is clear that ${\bf v}_u\in {\mathcal
A}^\beta_\alpha(V_1,\ldots,V_{n-1})$ for any $u\in L$ and, since
${\mathcal A}^\beta_\alpha$ is closed in ${\mathrm A}^\beta_\alpha$, we
have ${\bf v}_u\in {\mathcal A}^\beta_\alpha(V_1,\ldots,V_{n-1})$ for
any $u\in \bar L={\mathcal A}^{\prime\beta}_\alpha(V_n)$.
The changes in the proof for the case $\alpha=0$ are obvious.
The lemma is proved.

Let $V_1,\ldots,V_n$ be nonempty open connected cones in
$\R^{k_1},\ldots,\R^{k_n}$ respectively and let $K_j=V_j^*$, $j=1,\ldots,n$.
The product ${\mathcal A}^\beta_\alpha(V_1)
\otimes_i\ldots\otimes_i {\mathcal A}^\beta_\alpha(V_n)$
is continuously embedded into
${\mathcal A}^\beta_\alpha(V_1,\ldots,V_{n})$
by of the ordinary identification
$$
(\mathbf v_1\otimes\ldots\otimes \mathbf v_n)(z_1,\ldots,z_n)=
\mathbf v_1(z_1)\ldots \mathbf v_n(z_n),\quad \mathbf v_j\in
{\mathcal A}^\beta_\alpha(V_j).
$$
We define the operator ${\mathcal L}_{V_1,\ldots,V_n}\,:\,
s^{\prime\alpha}_\beta(K_1,\ldots,K_n)\to
{\mathcal A}^\beta_\alpha(V_1,\ldots,V_{n})$ as the continuous extension
of
${\mathcal L}_{V_1}\otimes_i\ldots\otimes_i {\mathcal L}_{V_n}$
to $s^{\prime\alpha}_\beta(K_1,\ldots,K_n)$.
By Lemma~\ref{l5}
and in view of the completeness of ${\mathcal A}^\beta_\alpha$,
such an extension
exists and is uniquely defined.
For any $u\in s^{\prime\alpha}_\beta(K_1,\ldots,K_n)$, we have
\begin{equation}
({\mathcal L}_{V_1,\ldots,V_n}u)(z)=u(e^{i\dv{\cdot}{z}}),\quad z\in T^V,
\label{7}
\end{equation}
where $V=V_1\times\ldots\times V_n$.
Thus, ${\mathcal L}_{V_1,\ldots,V_n}$ is the restriction of the Laplace
operator ${\mathcal L}_{V_1\times\ldots\times V_n}$ to
$s^{\prime\alpha}_\beta(K_1,\ldots,K_n)$. To prove formula~(\ref{7}),
it suffices to
note that it holds for $u\in s^{\prime\alpha}_\beta(K_1)\otimes\ldots\otimes
s^{\prime\alpha}_\beta(K_n)$ and next to make use of Lemma~\ref{l5}
and the continuity of both sides of the equality in $u$.

\theorem
\label{t4}
{\it
Let $\beta>1$, $0\leq \alpha<1$, let $V_1,\ldots,V_n$ be nonempty
open connected cones in $\R^{k_1},\ldots,\R^{k_n}$ respectively, and let
$K_j=V_j^*$, $j=1,\ldots,n$. The Laplace transformation  ${\mathcal
L}_{V_1,\ldots,V_n}$ is a topological isomorphism of
$s^{\prime\alpha}_\beta(K_1,\ldots,K_n)$ onto ${\mathcal
A}^\beta_\alpha(V_1,\ldots,V_{n})$. If $u\in
s^{\prime\alpha}_\beta(K_1,\ldots,K_n)$, then $({\mathcal
L}_{V_1,\ldots,V_n}u)(\cdot+iy)$ tends to the Fourier transform of $u$ in
the strong topology of $S^{\prime\beta}_\alpha(\R^{k_1}\times\ldots\times
\R^{k_n})$ as $y\to 0$ inside any cone $V_1'\times\ldots\times V_n'$,
where $V_j'\Subset V_j$, $j=1,\ldots,n$.

Proof.} In~\cite{Sol2}
the statement was established for $n=1$ and it is sufficient to prove the
theorem supposing it holds for the spaces over $n-1$ cones.
The mapping ${\mathcal
L}_{V_1,\ldots,V_n}$ is injective as the restriction
of the injective operator
${\mathcal L}_{V_1\times\ldots\times V_n}$.
Let $\mathbf v\in {\mathcal A}^\beta_\alpha(V_1,\ldots,V_{n})$. We define
the bilinear form $b_{\mathbf v}$ on ${\mathcal
A}^{\prime\beta}_\alpha(V_1,\ldots,V_{n-1})\times{\mathcal
A}^{\prime\beta}_\alpha(V_{n})$ by $b_{\mathbf
v}(u_1,u_2)=u_1(\mathbf v_{u_2})$. By
Lemma~\ref{l6}, the form $b_{\mathbf v}$
is separately continuous. Let $T_1:s^{\alpha}_\beta(K_1,\ldots,K_{n-1})
\to {\mathcal
A}^{\prime\beta}_\alpha(V_1,\ldots,V_{n-1})$ ($T_2:
s^{\alpha}_\beta(K_{n})\to {\mathcal A}^{\prime\beta}_\alpha(V_{n})$)
be the dual\footnote{Since $s^\alpha_\beta$ are reflexive spaces,
we identify $s^{\prime\prime\alpha}_\beta$ with $s^\alpha_\beta$.}
mapping of ${\mathcal L}^{-1}_{V_1,\ldots,V_{n-1}}$ (resp., of ${\mathcal
L}^{-1}_{V_n}$).
By Lemma~\ref{l5}, the separately continuous bilinear form $B_{\mathbf
v}(f_1,f_2)= b_{\mathbf v}(T_1f_1,T_2f_2)$ on
$s^{\alpha}_\beta(K_1,\ldots,K_{n-1})\times s^{\alpha}_\beta(K_{n})$
uniquely determines a functional $u\in
s^{\prime\alpha}_\beta(K_1,\ldots,K_{n})$ such that $u(f_1\otimes
f_2)=B_{\mathbf v}(f_1,f_2)$. If $z=(z_1,\ldots,z_n)$, $\tilde z=
(z_1,\ldots,z_{n-1})$, and $z_j\in T^{V_j}$, $j=1,\ldots,n$, then
$$
u(e^{\dv{\cdot}{z}})=B_{\mathbf v}(e^{\dv{\cdot}{\tilde z}'},
e^{\dv{\cdot}{z_n}_n}) =
b_{\mathbf v}(\delta_{\tilde z},\delta_{z_n})=\mathbf v(z),
$$
where $\dv{p}{\tilde z}'=\sum_{j=1}^{n-1}\dv{p_j}{z_j}_j$, $p\in
\R^{k_1}\times\ldots\times\R^{k_{n-1}}$. Thus, $\mathbf v$ is
the Laplace transform of $u$, i.e., the operator ${\mathcal
L}_{V_1,\ldots,V_{n}}$ is bijective.
The open mapping theorem shows that ${\mathcal
L}_{V_1,\ldots,V_{n}}$ is a topological isomorphism.
If $u\in
s^{\prime\alpha}_\beta(K_1,\ldots,K_{n})$ and
$f\in
S^\beta_\alpha(\R^{k_1+\ldots+k_n})$, then
\begin{equation}
\int ({\mathcal
L}_{V_1,\ldots,V_{n}}u)(x+iy)\,f(x)\,\di x=
u(e^{-\dv{\cdot}{y}}\hat f),
\quad y\in V_1\times\ldots\times V_n.
\label{8}
\end{equation}
Indeed, the formula holds for $n=1$, see~\cite{Sol2}, and ${\mathcal
L}_{V_1,\ldots,V_{n}}u$ coincides with ${\mathcal
L}_{V_1\times\ldots\times V_{n}}u$.
The direct check shows that
$e^{-\dv{\cdot}{y}}\hat f\to \hat f$ in the topology of
$s^\alpha_\beta(K_1,\ldots,K_n)$ as
$y\to 0$ inside $V'_1\times\ldots\times V'_n$. Therefore,
to prove the last statement of the theorem, it suffices to apply~(\ref{8})
and to take into account that in the Montel space $S^{\prime\beta}_\alpha$, weak
convergence and strong convergence are equivalent.

\section{Euclidean reconstruction theorem}
\label{c1s5}

From now on, the Lorentz product $p^0 x^0-p^1 x^1-\ldots-p^{d-1}
x^{d-1}$ of $p,x\in\R^d$ will be denoted by $px$.

All requirements of the Wightman formalism except for the spectral
condition are formulated in the usual way for the fields of the class
$S^{\prime \beta}_\alpha$, $0\leq\alpha<1,\,\beta>1$ (under the condition
$\beta>1$, the local
commutativity is formulated as usual). As we have already noted in Introduction, the spectral
condition in standard form~(\ref{i2})
is inapplicable in this case because of the lack of test functions of compact
support
in $p$-space. To obtain an appropriate generalization of the spectral
condition, one can use the notion of strong carrier cone
introduced in Section~2. As a result, we come to the following set of
axioms for the Wightman functions:
\begin{itemize}
\item[W1] (Growth and singularity) $w_n\in S^{\prime\beta}_\alpha(\R^{dn})$
$0\leq\alpha<1,\,\beta>1$;
\item[W2] (Relativistic invariance) $w_n(\Lambda
x_1+a,\ldots,\Lambda x_n+a)= w_n(x_1,\ldots,x_n)$ for any proper
Lorentz transformation $\Lambda$ and vector $a\in \R^d$;
\item[W3]
(Generalized spectral condition) $\bar \V_+^n$ is a strong carrier cone of
$\hat W_n$, i.e.,  $\hat W_n\in
s^{\prime\alpha}_\beta(\bar\V_+,\ldots,\bar \V_+)$;
\item[W4] (Locality)
$w_n(x_1,\ldots,x_j,x_{j+1},x_n)-w_n(x_1,\ldots,x_{j+1},x_j,\ldots,x_n)=0$ if
$x_j-x_{j+1}$ is space-like.
\end{itemize}
We do not impose the positivity condition on $w_n$,
which corresponds to the case of an indefinite metric in the state space.
Besides,
we do not require the fulfillment of the cluster property which is not
equivalent to the uniqueness of the vacuum in indefinite metric theories,
see~\cite{MorStr}.

It should be noted that in the indefinite metric case, theory is not
determined uniquely by its Wightman functions and to obtain its complete
operator realization, it is necessary to specify, in addition
to the sequence $w_n$, the Hilbert majorant of the indefinite metric which
determines the convergence in the state space~\cite{MorStr}.
For simplicity, we restrict our consideration to Wightman functions and do
not touch here more subtle questions concerning the construction of the Hilbert
majorant.

Using Theorem~\ref{t4} and condition W3, we conclude\footnote{Here and
subsequently, applying
Theorem~\ref{t4}, we set $\dv{p}{x}=-\sum_{j=1}^n p_j x_j,\, p,x\in
\R^{dn}$.} that $W_n(\xi)$ is the boundary value of
the function $\mathbf W_n(\zeta)=(2\pi)^{-dn}{\mathcal
L}_{\V_-,\ldots,\V_-}\hat W_n$
holomorphic in the past tube $\R^{dn}+i\V_-^n$. Correspondingly, $w_n$
is the boundary value of the function
$\mathbf
w_n(z_1,\ldots,z_n)=\mathbf W_{n-1}(z_1-z_2,\ldots,z_{n-1}-z_n)$
holomorphic in the domain $\{z\,:\, z_j-z_{j+1}\in \R^d+i\V_-\}$.
Standard analysis~\cite{Jost} based on the relativistic invariance
and locality shows that $\mathbf
w_n$ can be continued analytically to the extended domain $O_n^{ext}$
which is invariant under the complex Lorentz transformations and
the permutations of arguments.
For $x=(x_1,\ldots,x_n)\in \R^{dn}$, we set $\iota
x=(\iota x_1,\ldots,\iota x_n)$, where $\iota
x_j=(ix_j^0,x_j^1,\ldots,x_j^{d-1})$.
Then
$\iota x\in O_n^{ext}$ if and only if $x\in
\R^{dn}_{\ne}=\{x\in \R^{dn}\,:\,x_i\ne x_j,\,1\leq i<j\leq n\}$,
see~\cite{OS1,BLOT}. The Schwinger functions
$s_n$ are defined by the relation $s_n(x)=\mathbf w_n(\iota x)$,
$x\in \R^{dn}_{\ne}$. In the same way as in the ordinary
theory~\cite{OS1,BLOT}, we establish that $s_n$ are rotationally
invariant and symmetric with respect to the permutations of arguments.
Let $S_n(\xi)$ be the Schwinger function
$s_{n+1}$ expressed in terms of the difference variables
$\xi_j=x_j-x_{j+1}$, and let $\R^{dn}_-=\{x\in
\R^{dn}\,:\,x_j^0<0,\,j=1,\ldots,n\}$. If $\xi\in \R^{dn}_-$, then $\iota
\xi$ lies in the past tube and by Theorem~\ref{t4}, the function
$S_n(\xi)=\mathbf W_n(\iota\xi)$ satisfies, for $0<\alpha<1$ (for
$\alpha=0$), the bound
\begin{align} & |S_n(\xi)|\leq
C_{\epsilon}\exp[\epsilon |\xi|^{1/\alpha}+\epsilon (\min\nolimits_{1\leq
j\leq n} |\xi_j^0|)^{-1/(\beta-1)}],\quad \xi\in \R^{dn}_-, \label{10}\\
&(\mbox{resp., } |S_n(\xi)|\leq C_{\epsilon,R} \exp[\epsilon
(\min\nolimits_{1\leq j\leq n} |\xi_j^0|)^{-1/(\beta-1)}], \quad \xi\in
\R^{dn}_-,\,|\xi|\leq R)
\nonumber
\end{align}
for any $\epsilon>0$ (resp.,
for any $\epsilon,R>0$). As shown in~\cite{BLOT} (see the proof of
Theorem 9.30), for any $x\in \R^{dn}_{\ne}$ there exist a rotation $T$ and
a permutation $\pi$ of the set $[1..n]$ such that
\begin{equation}
\min\nolimits_{1\leq j\leq n-1}[(Tx_{\pi(j+1)})^0-(Tx_{\pi(j)})^0] \geq c
\min\nolimits_{j\ne k}|x_j-x_k|,
\label{11}
\end{equation}
where $c$ is a positive constant depending only on $n$.
In view of the invariance of the Schwinger functions under rotations and
permutations of arguments, (\ref{10}) and~(\ref{11}) imply
the inequality
\begin{align}
& |s_n(x)|\leq C_{\epsilon}\exp[\epsilon |x|^{1/\alpha}+\epsilon
(\min\nolimits_{j\ne k}|x_j-x_k|)^{-1/(\beta-1)}],\quad x\in
\R^{dn}_{\ne}, \label{12}\\
& (\mbox{resp., } |s_n(x)|\leq
C_{\epsilon,R} \exp[\epsilon (\min\nolimits_{j\ne
k}|x_j-x_k|)^{-1/(\beta-1)}], \quad x\in
\R^{dn}_{\ne},\,|x|\leq R). \nonumber
\end{align}
for any $\epsilon>0$
(resp., for any $\epsilon,R>0$).
The obtained estimates allow interpreting $s_n$ as generalized functions
defined under smearing with suitable test functions. The relevant
test function spaces are introduced by the following definition.

\defin
\label{d4}
Let $\alpha\geq 0$, $\beta>1$ and let $O$ be an open set in
$\R^k$. We denote by
${\mathit\Sigma}^\beta_\alpha(O)$ the subspace of $S^\beta_\alpha(\R^k)$
consisting of those functions that are identically zero on the complement
$\complement O$ of $O$ together with all
their derivatives.

${\mathit\Sigma}^\beta_\alpha(O)$ is a closed subspace of
$S^\beta_\alpha(\R^k)$. Therefore, by Theorem~$7'$ of~\cite{Komatsu},
we have
${\mathit\Sigma}^\beta_\alpha(O)=\varinjlim_{A,B>0}
{\mathit\Sigma}^{\beta,B}_{\alpha,A}(O)$, where
${\mathit\Sigma}^{\beta,B}_{\alpha,A}(O)$ is the Banach space
consisting of the functions $f\in {\mathit\Sigma}^\beta_\alpha(O)$
such that $|||f|||_{B,A}<\infty$ (see formula~(\ref{0})).

\lemma
\label{l6a}
{\it Let $O$ be an open set in
$\R^k$. If $\alpha>0$
(if $\alpha=0$), then for any $A,B>0$ there is
$A'>0$ such that for all
$x\in O$ and $f\in {\mathit\Sigma}^{\beta,B}_{\alpha,A}(O)$
the inequality
\begin{align}
&|f(x)|\leq C |||f|||_{B,A}\exp[-A'|x|^{1/\alpha}-A'(\delta_{\complement
O}(x))^{-1/(\beta-1)}]
\nonumber \\
&(\mbox{resp., }|f(x)|\leq
C |||f|||_{B,A}
\exp[-A'(\delta_{\complement O}(x))^{-1/(\beta-1)}] \mbox{ and }
f(x)=0 \mbox{ for } |x|\geq A),
\nonumber
\end{align}
is valid, where $\delta_{\complement O}(x)$ is the distance from $x$ to
$\complement O$.

Proof.}
Let $f\in
{\mathit\Sigma}^{\beta,B}_{\alpha,A}(O)$, $x\in O$ and $x_0$ be a point
in $\complement O$ such that $|x-x_0|=\delta_{\complement O}(x)$.
By Taylor's formula, for every $m\in \N=0,1,\ldots$ we have
$f(x)=\sum_{|\lambda|=m}\partial^\lambda f(x_0+th)h^\lambda/\lambda!$,
where $0<t<1$, $h=x-x_0$, and the standard multi-index notation is used.
From~(\ref{0}) it follows that $|\partial^\lambda f(x)|\leq
|||f|||_{B,A}B^{|\lambda|}|\lambda|^{\beta|\lambda|}$. Since
$|h^\lambda|\leq |h|^{|\lambda|}$, we get $|f(x)|\leq |||f|||_{B,A}
(B|h|)^m m^{\beta m}\sum_{|\lambda|=m}1/\lambda!= |||f|||_{B,A} (B|h|k)^m
m^{\beta m}/m!$ and using the inequality $m!\geq (m/e)^m$, we find that
$|f(x)|\leq |||f|||_{B,A} \inf_{m\in\N}(B|h|k e)^m m^{(\beta-1)m}$.
As shown in~\cite{GS}, Sec. IV.2,
$\inf_m \xi^{-m} m^{\alpha m}\leq
\exp(-\frac{\alpha}{e}\xi^{1/\alpha}+\alpha e/2)$ for any $\xi,\alpha>0$.
Replacing $\alpha$ and $\xi$ with $\beta-1$ and $1/B|h|ke$ respectively,
we obtain
\begin{equation}
|f(x)|\leq C_1 |||f|||_{B,A} \exp\left(-\frac
{(\beta-1)}{e}(Bke\delta_{\complement
O}(x))^{-\frac{1}{\beta-1}}\right).
\label{13}
\end{equation}
On the other hand, by~(\ref{0}) we have $|f(x)|\leq |||f|||_{B,A}
\inf_{m\in \N}(A/|x|)^m m^{\alpha m}$. For $\alpha=0$, this implies that
$f(x)=0$ for $|x|>A$. If $\alpha>0$, then an analogous estimation of the infimum shows
that $|f(x)|\leq
C_2 |||f|||_{B,A}\exp(-\frac{\alpha}{e}(|x|/A)^{1/\alpha})$. Multiplying
the last estimate and inequality~(\ref{13}) and taking the square root
of the left- and right-hand sides, we arrive at the statement of the lemma.

Since $\delta_{\complement \R^{dn}_{\ne}}(x)\leq
\min\nolimits_{j\ne k}|x_j-x_k|$ for $x\in \R^{dn}_{\ne}$,
Lemma~\ref{l6a} and the estimate~(\ref{12}) imply that $s_n\in
{\mathit\Sigma}^{\prime\beta}_\alpha(\R^{dn}_{\ne})$. Analogously,
from~(\ref{10}) it follows that $S_n\in
{\mathit\Sigma}^{\prime\beta}_\alpha(\R^{dn}_-)$.

For $\mathbf v\in {\mathcal A}^{\beta}_\alpha(\V_-,\ldots,\V_-)$, we set
$l_{\mathbf v}(f)=(2\pi)^{-dn}\int_{\R^{dn}_-}\mathbf v(\iota x) f(x)\,\di
x$, $f\in {\mathit\Sigma}^{\beta}_\alpha(\R^{dn}_-)$. By
Lemma~\ref{l6a}, the mapping $\mathbf v\to l_{\mathbf v}$
from ${\mathcal
A}^{\beta}_\alpha(\V_-,\ldots,\V_-)$ into
${\mathit\Sigma}^{\prime\beta}_\alpha(\R^{dn}_-)$ is
continuous. Consequently,
for every fixed $f\in {\mathit\Sigma}^{\beta}_\alpha(\R^{dn}_-)$ the functional
$u \to l_{{\mathcal L}_{\V_-,\ldots,V_-}u}(f)$ is continuous on
$s^{\prime\alpha}_\beta(\bar\V_+,\ldots,\bar\V_+)$ and because of
the reflexivity of the latter space there is an element
$\check f\in
s^{\alpha}_\beta(\bar\V_+,\ldots,\bar\V_+)$ such that
\begin{equation}
(2\pi)^{-dn}\int_{\R^{dn}_-} ({\mathcal L}_{\V_-,\ldots,V_-}u)(\iota
\xi)f(\xi)\,\di \xi= u(\check f),\quad u\in
s^{\prime\alpha}_\beta(\bar\V_+,\ldots,\bar\V_+).
\label{14}
\end{equation}
Taking $u=\delta_p$ (the value of $\delta_p$ on a test function $g$
is equal to $g(p)$),
we find that
\begin{equation}
\check f(p) = (2\pi)^{-dn}\int_{\R^{dn}_-}
f(\xi)\exp\left[\sum\nolimits_{j=1}^n (p_j^0 \xi_j^0-
i p^1_j\xi^1_j-\ldots-i p^{d-1}_j\xi^{d-1}_j)\right]\,\di
\xi.
\label{15}
\end{equation}
The mapping $f\to \check f$
from ${\mathit\Sigma}^{\beta}_\alpha(\R^{dn}_-)$ to
$s^{\alpha}_\beta(\bar\V_+,\ldots,\bar\V_+)$
has the continuous injective mapping
$u\to
l_{{\mathcal L}_{\V_-,\ldots,V_-}u}$ as its dual.
As a consequence, it is a continuous mapping with dense image.

\lemma
\label{l6b}
{\it The mapping $f\to\check f$ defined by $(\ref{15})$
is a continuous dense embedding of
${\mathit\Sigma}^{\beta}_\alpha(\R^{dn}_-)$ into
$S^\alpha_\beta(\R^{dn}_+)$, where $\R^{dn}_+=-\R^{dn}_-$.}

To prove the lemma, we need the following auxiliary statement.

\lemma
\label{l6c}
{\it Let $V_1$ and $V_2$ be nonempty open convex cones in
$\R^{k_1}$ and $\R^{k_2}$ respectively. Then
${\mathit\Sigma}^{\beta}_\alpha(V_1\times
V_2)={\mathit\Sigma}^{\beta}_\alpha(V_1)\hat\otimes_i
{\mathit\Sigma}^{\beta}_\alpha(V_2)$.

Proof.}
Applying Lemma~\ref{l4} to the nuclear DFS-spaces
${\mathit\Sigma}^{\beta}_\alpha(V_{1,2})$, we obtain
${\mathit\Sigma}^{\beta}_\alpha(V_1)\otimes_i
{\mathit\Sigma}^{\beta}_\alpha(V_2)={\mathit\Sigma}^{\beta}_\alpha(V_1)\otimes_e
{\mathit\Sigma}^{\beta}_\alpha(V_2)$ and by Lemma~\ref{l5a},
it suffices to show that the tensor product
${\mathit\Sigma}^{\beta}_\alpha(V_1)\otimes {\mathit\Sigma}^{\beta}_\alpha(V_2)$
is dense in ${\mathit\Sigma}^{\beta}_\alpha(V_1\times V_2)$.
In other words, we have to demonstrate that if a functional $u\in
S^{\prime\beta}_\alpha(\R^{k_1+k_2})$ vanishes on
${\mathit\Sigma}^{\beta}_\alpha(V_1)\otimes
{\mathit\Sigma}^{\beta}_\alpha(V_2)$, then it also vanishes on
${\mathit\Sigma}^{\beta}_\alpha(V_1\times V_2)$.
To this end, we take
$\psi_{1,2}\in {\mathit\Sigma}^{\beta}_\alpha(-V_{1,2})$ such that
$\int_{\R^{k_{1,2}}}\psi_{1,2}\,\di x=1$ and set
$\Psi_\varepsilon(x_1,x_2)=\varepsilon^{-k_1-k_2}\psi_1(x_1/\varepsilon)
\psi_2(x_2/\varepsilon)$. If $x\in\bar V_1\times\bar V_2$, then
$\Psi_\varepsilon(x-\cdot)\in {\mathit\Sigma}^{\beta}_\alpha(V_1)\otimes
{\mathit\Sigma}^{\beta}_\alpha(V_2)$ and, consequently,
$(u*\Psi_\varepsilon)(x)=0$. Hence, for $f\in
{\mathit\Sigma}^{\beta}_\alpha(V_1\times V_2)$, we have
$u(f)=\lim_{\varepsilon\to 0}\int_{\bar V_1\times\bar V_2}
(u*\Psi_\varepsilon)(x) f(x)\,\di x=0$. The lemma is proved.

{\it Proof of Lemma~$\ref{l6b}$.}
If $\check f=0$, then setting
$u=\delta_{\iota p}$ in~(\ref{14}), we see that the Fourier transform
of
$f$ vanishes and hence $f=0$. Thus, the mapping $f\to
\check f$ is injective.
For $f\in \Sigma^\beta_\alpha(\R_-)$,
we set
$\tilde f(p) = (2\pi)^{-1}\int_{\R_-} f(\xi) e^{\xi p} \di\xi$.
In the same way as above
(see the paragraph preceding the formulation of Lemma~\ref{l6b}),
we establish that $\tilde f\in
s^{\alpha}_{\beta}(\bar\R_+)=S^{\alpha}_{\beta}(\R_+)$ and that the mapping
$P$ taking
$f$ to $\tilde f$ is a continuous dense embedding
of $\Sigma^\beta_\alpha(\R_-)$ into $S^{\alpha}_{\beta}(\R_+)$.
By Lemma~\ref{l6c}, we have $\Sigma^\beta_\alpha(\R^d_-)=
\Sigma^\beta_\alpha(\R_-)\hat\otimes_i \Sigma^\beta_\alpha(\R^{d-1})$ and
Theorem~3 of~\cite{Sol2} ensures that $S^{\alpha}_{\beta}(\R^d_+)=
S^{\alpha}_{\beta}(\R_+)\hat\otimes_i S^{\alpha}_{\beta}(\R^{d-1})$.
Let
$L_1=P\hat\otimes_i {\mathcal F}$, where
$\mathcal F$ is the (inverse) Fourier transformation on $\R^{d-1}$:
$$
({\mathcal F}f)(p^1,\ldots,p^{d-1})= (2\pi)^{-(d-1)}\int_{\R^{d-1}}
f(\xi)e^{-i\xi^1 p^1-\ldots-i\xi^{d-1}p^{d-1}}\di\xi.
$$
Obviously, $L_1$ is a continuous operator from
$\Sigma^\beta_\alpha(\R^d_-)$ to $S^{\alpha}_{\beta}(\R^d_+)$ with a
dense image. Besides, $(L_1 f)(p)=\check f(p)$ for all $f\in
\Sigma^\beta_\alpha(\R^d_-)$. Indeed, this equality holds for
$f\in \Sigma^\beta_\alpha(\R_-)\otimes \Sigma^\beta_\alpha(\R^{d-1})$, and
since both sides of the equality are continuous in $f$, it is valid
everywhere on $\Sigma^\beta_\alpha(\R^d_-)$. Thus, the lemma is proved for
$n=1$.  For $n>1$, we make use of the representations
$\Sigma^\beta_\alpha(\R_-^{dn})=
\Sigma^\beta_\alpha(\R^d_-)\hat\otimes_i\ldots\hat\otimes_i
\Sigma^\beta_\alpha(\R^d_-)$ and
$S^\alpha_\beta(\R_+^{dn})=S^\alpha_\beta(\R^d_+)
\hat\otimes_i\ldots\hat\otimes_iS^\alpha_\beta(\R^d_+)$
which follow by induction from~(\ref{6}), Lemma~\ref{l6c} and
Theorem~3 of~\cite{Sol2}.
Setting
$L_n=L_1\hat\otimes_i\ldots\hat\otimes_i L_1$ and arguing as above,
we make sure that $L_n$ is a continuous operator from
$\Sigma^\beta_\alpha(\R^{dn}_-)$ to $S^\alpha_\beta(\R^{dn}_+)$ with dense
image and such that $(L_n f)(p)=\check f(p)$. The lemma is proved.

Substituting $u=\hat W_n$ in~(\ref{14}) yields
\begin{equation}
\int_{\R^{dn}_-} S_n(x)f(x)\,\di x= \hat W_n(\check f),\quad
f\in {\mathit\Sigma}^{\beta}_\alpha(\R^{dn}_-).
\label{16}
\end{equation}
By condition (W3), there is a continuous seminorm $P$ on
$S^\alpha_\beta(\R^{dn}_+)$ such that $|\hat W_n(f)|\leq P(f)$ for every
test function in $S^\alpha_\beta(\R^{dn}_+)$. By Lemma~\ref{l6b} and
equality~(\ref{16}), it hence follows that $|S_n(f)|\leq P(\check f)$,
$f\in {\mathit\Sigma}^{\beta}_\alpha(\R^{dn}_-)$.

Summarizing the above discussion, we obtain the following set of conditions
on the Schwinger functions:
\begin{itemize}
\item[S1] (Growth and singularity) $s_n\in
{\mathit\Sigma}^{\prime\beta}_\alpha(\R^{dn}_{\ne})$;
\item[S2] (Euclidean invariance) $s_n(T x_1+a,\ldots,T
x_n+a)= s_n(x_1,\ldots,x_n)$ for any rotation
$T$ and any $a\in \R^d$;
\item[S3] (Laplace transform condition) There is a continuous seminorm
$P$ on $S^\alpha_\beta(\R^{dn}_+)$ such that for every $f\in
{\mathit\Sigma}^{\beta}_\alpha(\R^{dn}_-)$ the inequality
$|S_n(f)|\leq P(\check f)$ holds, where $\check f$ is the function defined
by formula~(\ref{15});
\item[S4] (Symmetry)
$s_n(x_{\pi(1)},\ldots,x_{\pi(n)})=s_n(x_1,\ldots,x_n)$ for all
permutations $\pi$ of the indices.
\end{itemize}

We now can formulate the main result.
\theorem
\label{t5}
{\it
For a given sequence of Wightman functions
$w_n$ satisfying
{\rm W1--W4}, the corresponding sequence of the Schwinger functions $s_n$
satisfies {\rm S1--S4}. Conversely, generalized functions
satisfying {\rm S1--S4} are the Schwinger functions corresponding
to a uniquely determined sequence of Wightman functions satisfying
{\rm W1--W4}.

Proof. }
The construction of the Schwinger functions corresponding to given Wightman
functions
and the derivation of the properties {\rm S1--S4} are given
above and we we only need to prove the converse statement. Let the sequence $s_n$
satisfy S1--S4 and let $L$ denote the image of $\R^{dn}_-$ under
the mapping $f\to \check f$. By S3, the linear functional $\check f\to
S_n(f)$ defined on $L$ is continuous in the topology of
$S^{\alpha}_{\beta}(\R^{dn}_+)$ and in view of Lemma~\ref{l6b}
there is a uniquely determined generalized function
$\hat W_n\in
S^{\prime\alpha}_{\beta}(\R_+^{dn})$ such that $\hat W_n(\check
f)=S_n(f)$, $f\in \Sigma_\alpha^\beta(\R^{dn}_-)$.  The invariance of $\hat
W_n$ under spatial rotations follows immediately from
S2. To prove the invariance of $\hat W_n$ under pure Lorentz
transformations, it suffices to show that
$X_{0l} \hat W_n=0$, where
$l=1,2,3,\,\,X_{0l}=\sum_{k=0}^n (p_k^0\partial/\partial p_k^l+
p_k^l\partial/\partial p_k^0)$ are the infinitesimal generators of boosts.
Let $Y_{0l}=\sum_{k=0}^n (\xi_k^0\partial/\partial \xi_k^l-
\xi_k^l\partial/\partial \xi_k^0)$ be the infinitesimal generators of
Euclidean rotations. It is easy to see that $Y_{0l}f$
is taken to
$X_{0l}\check f$ by the mapping $f\to \check f$ and hence
$X_{0l}\hat W_n$ vanishes on $L$:
$$ (X_{0l}\hat
W_n)(\check f)=-W_n(X_{0l}\check f)= -S_n(Y_{0l}f)=(Y_{0l}S_n)(f)=0,\quad
f\in \Sigma_\alpha^\beta(\R^{dn}_-).  $$
Using Lemma~\ref{l6b} and the continuity of
$\hat W_n$, we conclude that $X_{0l}\hat W_n=0$. By
the proven Lorentz invariance, $\hat W_n$ belongs not only to
$S^{\prime\alpha}_{\beta}(\R^{dn}_+)$, but also to every space
$S^{\prime\alpha}_{\beta}((\Lambda \R^d_+)^n)$, where $\Lambda$ is a
proper Lorentz transformation, and, moreover, to every space
$s^{\prime\alpha}_{\beta}(\Lambda \bar \R^d_+,\ldots,\Lambda \bar\R^d_+)$.
Applying Theorem~\ref{t2} and using the equality $\cap_\Lambda \Lambda \bar
\R^d_+=\bar \V_+$, we conclude that $\bar \V_+^n$ is a strong carrier cone
of $\hat W_n$. We now define the Wightman functions $w_n$
by formula~(\ref{i3}) and the second relation in~(\ref{i2}). Obviously,
$w_n$ satisfy conditions W1, W2, and W3. Substituting $u=\hat W_n$
in~(\ref{14}) shows that $s_n$ are indeed the Schwinger functions
corresponding to $w_n$. The symmetry of $s_n$ implies the symmetry of
the Wightman functions $w_n$ in their ordinary analyticity domain, whence
property W4 is derived by the standard arguments~\cite{Jost}.  The theorem
is proved.

\section{Wick power series}
\label{c1s6}

In this section, we show that the generalized spectral
condition formulated in the previous section is satisfied for the simplest examples of quantum fields with
highly singular infrared behavior, namely, for the fields representable by
infinite series in the Wick powers of an indefinite metric free field $\phi$, i.e., by series
of the form
\begin{equation}
\sum_{k=0}^{\infty} d_k \W{\phi^k}(x).
\label{17}
\end{equation}
We assume
that $\phi$
is a tempered operator-valued distribution acting in
a Hilbert-Krein state space $\mathcal H$ (see~\cite{MorStr} for the role of this
condition). This means that $\mathcal H$
is endowed, in addition to an indefinite metric $\dv{\cdot}{\cdot}$, by an
auxiliary positive scalar product $(\cdot,\cdot)$ connected with
$\dv{\cdot}{\cdot}$ by the relation $\dv{\Phi}{\Psi}=(\Phi,\theta \Psi)$,
where $\Phi,\Psi\in \mathcal H$ and $\theta$ is a bounded self-adjoint
operator such that $\theta^2=1$. The scalar product $(\cdot,\cdot)$ determines a
distribution $\maj$, which is called the majorant of the two-point vacuum
average $w(x-x')=\dv{\Psi_0}{\phi(x)\phi(x')\Psi_0}$, by the relation
\begin{equation}
(\phi(f)\Psi_0,\phi(g)\Psi_0)=\int\maj(x,x')\bar f(x) g(x')\,\di x\di x',
\nonumber
%\label{}
\end{equation}
where $\Psi_0$ is the vacuum and $f,g$ are test functions in the Schwartz
space $S(\R^d)$. The the Krein structure implies~\cite{SS1}
that $\maj(x,x')$ is the boundary value of a function $\mjz(z,z')$
holomorphic in the tubular domain $\{\,(z,z')\in \C^{2d}:
y=\mathrm{Im}\,z\in \V_-,\,y'=\mathrm{Im}\,z'\in \V_+\}$.
As in~\cite{SS2}, we find it convenient
to characterize the infrared and ultraviolet behavior of the majorant by a
pair of monotone nonnegative functions $\wir$ and $\wuv$
increasing as their arguments tend to infinity and to zero,
respectively, and satisfying the estimate
\begin{equation}
|\mjz(z,z')|\leq C(1 + \wir(|z|+|z'|)+\wuv(|y|+|y'|)),\quad
(y,y')\in V\times V',
\label{18}
\end{equation}
for any compact subcones $V$ and $V'$ of $\V_-$ and $\V_+$ (with constant
$C$ depending on $V$ and $V'$).
Formula~(\ref{18}) also allows to estimate the analytic two-point Wightman
function $\mathbf{w}(z)$ because
\begin{equation}
|{\bf w}(x-x'-2iy)|^2\leq |{\bf w}_{{\rm maj}}(x-iy,\, x+iy)|
\,|{\bf w}_{{\rm maj}}(x'-iy,\, x'+iy)|
\label{18a}
\end{equation}
for all $y\in \V_+$. Indeed, as $\theta^2=1$, we have
$$
|\langle\phi(f)\Psi_0,\phi(g)\Psi_0 \rangle|\leq \|\phi(f)\Psi_0\|
\|\phi(g)\Psi_0\|\,.
$$
Taking $f(\xi)=(\nu/\sqrt\pi)^{\tt d}e^{-\nu^2(\xi-x-iy)^2}$ and
$g(\xi)=(\nu/\sqrt\pi)^{\tt d}e^{-\nu^2(\xi-x'-iy)^2}$ and writing the
left- and right-hand sides in this inequality as integrals over a
 plane in the analyticity domain and passing to the limit as
 $\nu\to\infty$, we immediately obtain
 (\ref{18a}). Choosing $V'=-V$ in (\ref{18}) and substituting (\ref{18}) in (\ref{18a}) yield
\begin{equation}
|\mathbf w(\zeta)|\leq C(1 + \wir(2|\zeta|)+\wuv(|\eta|)),\quad
\eta=\mathrm{Im}\,\zeta\in V,
\label{18b}
\end{equation}
for any compact subcone $V$ of $\V_-$ with $C$ depending on $V$.

The following
criterion allows finding the adequate test function space on which the
series (\ref{17}) is convergent.
\theorem
\label{t6}
{\it Let $\phi$ be a free field acting in a
Hilbert-Krein space $\mathcal H$, and let the positive majorant of its
    two-point Wightman function satisfy the inequality $(\ref{18})$ with monotonic
    $w_{{\scriptscriptstyle IR}}$ and $w_{{\scriptscriptstyle UV}}$.
    Let the coefficients $d_k$ satisfy the condition
\begin{equation}\label{19}
    |d_kd_l| \le Ah^{k+l} |d_{k+l}|
\end{equation}
with some $A,h>0$.
Then the series $(\ref{17})$ is well defined
   as an operator-valued generalized function on every space $S^\beta_\alpha$
    such that $\alpha>0$, $\beta>1$, and
    the relations
    \begin{equation}
    \sum_k L^kk!|d_{2k}| w_{{\scriptscriptstyle IR}}(r)^k \le
      C_{L,\epsilon}\,e^{\epsilon r^{1/\alpha}},\quad
      \sum_k L^kk! |d_{2k}| w_{{\scriptscriptstyle UV}}(t)^k \le
      C_{L,\epsilon}\, e^{\epsilon t^{-1/(\beta-1)}}
   \label{20}
   \end{equation}
hold for an arbitrarily large $L>0$ and an arbitrarily small $\epsilon>0$.}

This theorem follows immediately from Theorem~3 of~\cite{SS2} because (\ref{20}) implies
the inequality
$$
\inf_{t>0} e^{st}\sum_k L^kk! |d_{2k}| w_{{\scriptscriptstyle UV}}(t)^k \le
      C_{L,\epsilon}\, \exp[\beta(\epsilon/(\beta-1))^{(\beta-1)/\beta} s^{1/\beta}].
$$
It is straightforward to verify that in the case $\alpha>1$, the sum of the series~(\ref{17})
satisfies the usual Wightman axioms (except positivity). For $0<\alpha<1$, we have
the following theorem strengthening the results of~\cite{SS}.
\theorem
\label{t7}
{\it Under the conditions of Theorem~$\ref{t6}$, the Wightman functions of the field
$\varphi(x)=\sum_{k=0}^{\infty} d_k \W{\phi^k}(x)$ satisfy the requirements {\rm W1--W4}
including the generalized spectral condition.

Proof.} The only nontrivial point is to check the fulfilment of the generalized spectral condition.
The expression for the $n$-point vacuum expectation value of the field
$\varphi$ given by the Wick theorem is a power series in
$n(n-1)/2$ variables $w(x_j-x_m)$ and can be written as
   \begin{equation}
   \langle
     \Psi_0,\, \varphi(x_1)\ldots \varphi(x_n)\Psi_0\rangle
     =\sum_KD_K\,w^K\,,
   \nonumber
   \end{equation}
where $K$ is an integer-valued vector with nonnegative components
  $k_{jm}$, $1\leq j<m\leq n$, and $w^K(x)$ is the boundary value of the function
$
   \mathbf w^K(z)
=\prod_{j<m}\mathbf w(z_j-z_m)^{k_{jm}}
$
analytic in the tubular domain $\{\, z\in \C^{dn}: z_j-z_m\in \R^d+i\V_-,\,1\leq j<m\leq n\,\}$.
The usual combinatorial analysis related to the Wick theorem shows
that
 $$
  D_K={\kappa!\over K!}\prod_{1\leq j\leq n}d_{\kappa_j},
 $$
 where  $\kappa_j=k_{1j}+\ldots+k_{j-1,j}+k_{j,j+1}+\ldots+k_{jn}$ is
the total number of pairings in the given term of the series that involve
the argument
 $x_j$, and we follow the usual convention
$$
K!= \prod_{j<m}k_{jm}!\,,\quad
\kappa!= \prod_{1\leq j\leq n}\kappa_j!\,.
$$
Correspondingly, the $n$-point Wightman function expressed in terms of the difference variables is given by
   \begin{equation}
     W_{n-1}(\xi)=
     \sum_KD_K\,W^K(\xi)\,,
   \label{21}
   \end{equation}
where $W^K(\xi)$ is the boundary value of the function
$\mathbf W^K(\zeta)=\prod_{j<m}\mathbf w(\zeta_j+\ldots+\zeta_{m-1})^{k_{jm}}$ analytic in the
domain $\R^{d(n-1)}+i\V^{n-1}_-$. To prove the theorem, it is sufficient to establish that the series
$\sum_K D_K \mathbf W^K$ converges unconditionally in
${\mathcal A}^\beta_\alpha(\V_-,\ldots,V_-)$. Indeed, in this case, Theorem~\ref{t4} shows that
$\hat W_{n-1}$, which is the inverse Laplace transform of the sum of this series, belongs to
$s^{\prime \alpha}_{\beta}(\V_+,\ldots,\V_+)$, i.e., the generalized spectral condition is satisfied.
Since ${\mathcal A}^\beta_\alpha$ is complete, it suffices to verify that
\begin{equation}\label{22}
\sum_K |D_K| \|\mathbf{W}^K\|_{\epsilon,V_1,\ldots,V_{n-1}}<\infty
\end{equation}
for any $\epsilon>0$ and any cones $V_1,\ldots,V_{n-1}$ compact in $\V_-$. Let $V$ be the closed convex
hull of the union $V_1\cup\ldots\cup V_{n-1}$. The cone $V$
is the second
dual cone of $V_1\cup\ldots\cup V_{n-1}$ and, therefore,
is a compact subcone of $\V_-$ (because if $V\Subset U$ and $U$ is an open cone, then $U^*\subset \mathrm{int}\, V^*$).
Obviously, $\eta_j+\ldots+\eta_{m-1}\in V$ for any $\eta=(\eta_1,\ldots,\eta_{n-1})\in V_1\times
\ldots\times V_{n-1}$ and for any $1\leq j<m\leq n$. Further, there is a $\lambda>0$ such that
\begin{equation}\label{23}
    |\eta_j+\ldots+\eta_{m-1}|\geq \lambda(|\eta_j|+\ldots+|\eta_{m-1}|)
\end{equation}
for all $\eta\in \bar \V_-$ and $j<m$. Indeed, for fixed $j$ and $m$, (\ref{23}) is fulfilled if
we take
$$
\lambda=\lambda_{jm}=\inf_{(\eta_j,\ldots,\eta_{m-1})\in \bar \V_-^{m-j},\,|\eta_j|+\ldots+|\eta_{m-1}|=1}
|\eta_j+\ldots+\eta_{m-1}|.
$$
By the convexity of $\bar \V_-$, we have $\eta_j+\ldots+\eta_{m-1}=0$ if
and only if $\eta_j=\ldots=\eta_{m-1}=0$.
This implies that $\lambda_{jm}>0$ because the infimum is taken over a compact set.
So we can set $\lambda=\min_{j<m}\lambda_{jm}$. By (\ref{18b}), (\ref{23}), and the monotonicity of $\wir$ and
$\wuv$, for $\zeta\in \R^{d(n-1)}+iV_1\times\ldots\times V_{n-1}$ we have
\begin{equation}\label{24}
|\mathbf W^K(\zeta)|\leq (n+1)^{|K|}C^{|K|}(1+\wir(2n|\zeta|)^{|K|}+\sum\nolimits_{i=1}^{n-1}
\wuv(\lambda |\eta_i|)^{|K|}),
\end{equation}
where $|K|=\sum_{j<m} k_{jm}$. The condition (\ref{19}) and the inequalities $|K|!/K!\leq
(n(2n-1))^{|K|}$ and $\kappa!\leq|\kappa|!\leq 4^{|K|}(|K|!)^2$ following
from the well-known properties of polynomial coefficients yield
\begin{equation}
    |D_K|\leq A'h^{\prime |K|}|K|!\,|d_{2|K|}|,
   \label{25}
   \end{equation}
where the constant $h'$ depends on $n$. If $\wir$ and $\wuv$ are not both identically zero (which is assumed),
then (\ref{20}) implies that for any $L>0$, there is a $\tilde C_L$ such that
\begin{equation}\label{26}
k!|d_{2k}|\leq \tilde C_L L^{-k},\quad k=0,1\ldots
\end{equation}
Using (\ref{20}), (\ref{24}), (\ref{25}), (\ref{26}), and Definition~\ref{d3}, we obtain
$$
|D_K|\, \|\mathbf{W}^K\|_{\epsilon,V_1,\ldots,V_{n-1}}\leq
C'_{L,\epsilon}((n+1)Ch'/L)^{|K|}.
$$
This proves (\ref{22}) because the number of multi-indices $K$ with fixed $|K|$ depends polynomially
on $|K|$ and $L$ is arbitrarily large. The theorem is proved.

\section{Conclusion}
\label{c1s7}

We see that
the proposed formulation of
the spectral condition offers a means for a reasonable generalization
of a considerable part of the
Wightman-type formalism to
quantum fields with highly singular infrared behavior. In particular, gauge-dependent quark fields,
which were claimed in~\cite{RotheSchroer} to be ill-defined mathematical objects, can be
treated in this enlarged axiomatic framework.
This situation is somewhat analagous to that
in nonlocal QFT, where the corresponding generalization of local commutativity
ensures the preservation of the
PCT-symmetry~\cite{SolovievJMP} and the spin-statistics relation~\cite{Soloviev}, i.e.,
those basic physical results that are commonly believed to be consequences of locality.

In this paper, we have made no attempt to
derive an appropriate extension to infrared singular fields of the Osterwalder-Schrader linear
growth estimates which also ensure the reconstruction of Wightman functions from Schwinger functions and which
proved to be effective in constructive QFT. At first glance, there are no obstacles for obtaining such a
generalization provided the positivity condition is kept.
However, this condition is violated for all
relevant examples of infrared singular quantum fields. For this reason, we
confined our consideration to the indefinite metric case.

We conclude by a remark on how a notion analogous to that of a strong carrier cone of an analytic functional
can be introduced in the framework of Fourier hyperfunctions (i.e., functionals
defined on $S^{1}_1$) which is universal for local QFT~\cite{Nagamachi,BruningNagamachi}.
The construction given below is parallel to that of Section~\ref{c1s2}.

\textbf{Definition 1$'$.}
Let $U$ be an open set in $\R^k$.
The Banach space $S^{1,A}_{1,B}(U)$
consists of functions analytic in the $1/A$-neighborhood $U_{1/A}$ of $U$ in
$\C^k$ and having the finite norm
\begin{equation}
\|f\|_{U,A,B}= \sup_{w\in U_{1/A}} |f(w)|\exp(|p/B|).
\nonumber
\end{equation}
The space $S^{1}_{1}(U)$ is defined to be the inductive limit
$\varinjlim_{A,B>0}S^{1,A}_{1,B}(U)$.

Let $\mathcal R^k$ be the radial compactification of $\R^k$. For $\mathcal U\subset \mathcal R^k$,
we denote by $\mathcal{\tilde U}$ the set $\mathcal U\cap\R^k$. A compact set $\mathcal K\subset \mathcal R^k$
is said to be a carrier of a functional $u\in S^{\prime 1}_1(\R^k)$ if $u$ has a continuous extension to the space
$s^1_1(\mathcal K)= \varinjlim_{\mathcal U}S^{1}_{1}(\mathcal{\tilde U})$, where $\mathcal U$ runs
over all open neighborhoods of $\mathcal K$ in $\mathcal R^k$. The following definition is an analogue of
Definition~\ref{d2}.

\textbf{Definition 2$'$.}
Let $\mathcal K_1,\ldots,\mathcal K_n$ be compact sets in
$\mathcal R^{k_1},\ldots,\mathcal R^{k_n}$ respectively.
The functional $u\in S^{\prime
1}_{1}(\R^{k_1+\ldots+k_n})$ is said to be carried by the family of sets
$\mathcal K_1,\ldots,\mathcal K_n$
if $u$ has a continuous
extension to the space
$s^{1}_{1}(\mathcal K_1,\ldots,\mathcal K_n)=
\varinjlim_{\mathcal U_1,\ldots,\mathcal U_n}S^{1}_{1}
(\mathcal{\tilde U}_1\times\ldots\times \mathcal {\tilde U}_n)$,
where the inductive limit is taken over all open neighborhoods
$\mathcal U_1,\ldots,\mathcal U_n$ of the sets $\mathcal K_1,\ldots,\mathcal K_n$ respectively.

For $K\subset \R^k$, we denote by $\hat K$ the closure of $K$ in $\mathcal R^k$.
Let $K_1,\ldots K_n$ be closed sets in $\R^{k_1},\ldots,\R^{k_n}$ respectively, and let
$K=K_1\times\ldots\times K_n$.
The following example shows that the space $s^{\prime 1}_1(\hat K)$ is, in general,
different from $s^{\prime 1}_1(\hat K_1,\ldots,\hat K_n)$.

\textbf{Example 1$'$.} Let $k_1=k_2=1$, $K_1=\bar \R_+$, $K_2=\{0\}$, and $K=\bar\R_+\times\{0\}$.
Clearly, $s^{\prime 1}_1(\hat K_1,\hat K_2)\subset s^{\prime 1}_1(\hat K)$. In this case, we can assume
that $\mathcal {\tilde U}_{1,2}$ in Definition~2$'$ are just
the $\varepsilon$-neighborhoods of $\bar \R_+$ and $\{0\}$ respectively. If $\varepsilon<1/2$, then the sequence
$g_n(p)=p_2^n e^{-p_1}$ converges to zero in every space $S^{1,A}_{1,B}(\mathcal{\tilde U}_1\times
\mathcal{\tilde U}_2)$ with $A>2$ and $B>1$ and therefore is bounded in
$s^{1}_1(\hat K_1,\hat K_2)$. If the sequence $g_n$ were bounded in the DFS-space $s^1_1(\hat K)$, then it would be
bounded in some space $S^{1,A}_{1,B}(\mathcal{\tilde U})$, where $\mathcal U$ is an open neighborhood of $\hat K$.
However, any such $\mathcal U$ contains the ray
$r_\lambda=\{\,(p_1,p_2): p_1\geq 0,\, p_2=\lambda p_1\,\}$ with some $\lambda>0$ and, therefore, we have
$\|g_n\|_{\mathcal{\tilde U}, A,B}\geq \sup_{p\in r_\lambda}|g_n(p)|=\lambda^n n^n e^{-n}$.
Thus, the sequence $g_n$ is unbounded in $s^1_1(\hat K)$ and there is an $u\in s^{\prime 1}_1(\hat K)$
such that the number sequence $|u(g_n)|$ is unbounded (because any weakly bounded set in a locally
convex space is bounded). Obviously, $u$ does not belong to $s^{\prime 1}_1(\hat K_1,\hat K_2)$ and so
the latter is different from $s^{\prime 1}_1(\hat K)$.

This distinction may be essential for hyperfunction QFT, where the
spectral condition can be formulated in two alternative ways. One can require either that
$\hat W_n\in s^{\prime 1}_1(\hat\V_+^n)$
as in~\cite{Nagamachi,BruningNagamachi}
or that $\hat W_n\in s^{\prime 1}_1(\hat\V_+,\ldots,\hat\V_+)$.
It would be worthwhile to
examine the second condition from the viewpoint of the Euclidean formulation
of hyperfunction QFT, but this is beyond the scope of the present paper.

\section*{Acknowledgments}

The author
is grateful to M.~A.~Soloviev for helpful and stimulating discussions.
This research was supported by the Russian Foundation for Basic
Research (Grants No. 02-01-00556 and No. 00-15-96566) and INTAS (Grant No. 99-1-590).

\section*{Appendix~A. Proof of Lemma~\ref{l1} for $\alpha=0$.}

\lemma
\label{l7}
{\it Let $Q_1,Q_2$ be nonempty cones in $\R^k$ and
let $h_1,h_2$ be $C^\infty$-functions on
$\C^k$ such that $f=h_1+h_2$ is analytic in $\C^k$.
If the norms
$\|h_1\|_{Q_1,A,B}$, $\|h_2\|_{Q_2,A,B}$, and
$\|\partial h_1/\partial\bar w_j\|_{Q_1\cup Q_2,A,B}$ (given by~$(\ref{1})$
with $\alpha=0$) are finite for some $A,B>0$, then one can find $f_{1,2}\in S^0_\beta(Q_{1,2})$
such that $f=f_1+f_2$.}

Let us derive Lemma~\ref{l1} from
Lemma~\ref{l7}. Let $\chi\in C_0^\infty(\R^{k_1})$ satisfy
$\int_{\R^{k_1}}\chi(p')\di p'=1$. We set
$g_0(w')=\chi(\mathrm{Re}\,w')$ and define $g_{1,2}$ by
(\ref{1b}). The functions $g_{1,2}(w)$ as well as their derivatives
$\partial g_{1,2}/\partial \bar w_j$, $j=1,\ldots, k_1+k_2$,
satisfy the estimate $(\ref{2})$
with $\alpha=0$ for any $A_0,B_0>0$.
Therefore, setting $Q_{1,2}=(U\cup U_{1,2})\times V$ and repeating
the proof for nonzero $\alpha$, we conclude that
the norms $\|fg_2\|_{Q_1,A,B}$, $\|fg_1\|_{Q_2,A,B}$,
$\|f\partial g_{2}/\partial \bar w_j\|_{Q_1,A,B}$, and
$\|f\partial g_{1}/\partial \bar w_j\|_{Q_2,A,B}$ are finite for $A,B$
sufficiently large. Moreover, since $g_1+g_2=1$ and $\partial
g_{1}/\partial \bar w_j=- \partial g_{2}/\partial \bar w_j$, we have $\|f
\partial g_{2}/\partial \bar w_j\|_{Q_1\cup Q_2,A,B}<\infty$. Thus,
$h_{1,2}=f g_{2,1}$ satisfy the conditions of Lemma~\ref{l7}
because $\partial h_1/\partial \bar w_j=f\partial g_2/\partial \bar w_j$
in view of the analyticity of $f$. Lemma~\ref{l1}
is proved.

{\it Proof of Lemma~$\ref{l7}$}.
Let $h$ be a measurable function on $\C^k$ and $U$ be a nonempty cone in $\R^k$.
For
$a,b>0$ sufficiently large, from Definition~(\ref{1}) it follows that
\begin{equation}
C\|h\|_{U,A,B}\geq\|h\|'_{U,a,b}=
\left[\int_{\C^k}|h(w)|^2\exp(-\rho_{U,a,b}(w))\,\di\lambda(w)\right]^{1/2},
\label{a0}
\end{equation}
where
$\rho_{U,a,b}(p+iq)=-\sum_{j=1}^k |p_j/b|^{1/\beta}+
a\inf_{p'\in U}\sum_{j=1}^k |p_j-p'_j|+a\sum_{j=1}^k |q_j|$,
$\di\lambda$ is the Lebesgue measure on $\C^k$, and $C$ is a constant independent of
$h$.  If $h$ is analytic, then using Cauchy's integral
formula, one can prove~\cite{Sol1} the converse statement, i.e., if
$\|h\|'_{U,a,b}<\infty$ for some $a,b>0$, then $h\in S^0_{\beta}(U)$.

Suppose there is a locally integrable function $\psi$ on $\C^k$
which has the finite norm
$\|\psi\|'_{Q_1\cup Q_2,a_1,b_1}$ for
some $a_1,b_1>0$ and satisfies, as a generalized function,
the inhomogeneous Cauchy--Riemann equations
\begin{equation}
\frac{\partial\psi}{\partial \bar w_j} =\eta_j,\quad j=1,\ldots,k,
\label{a1}
\end{equation}
where $\eta_j=\partial h_1/\partial \bar w_j$. Then $f_1=h_1-\psi$ and
$f_2=h_2+\psi$ satisfy, as generalized functions, the homogeneous
equations $\partial f_{1,2}/\partial \bar w_j=0$ and, consequently,
are ordinary analytic functions. Moreover, in view of~(\ref{a0})
we have $\|f_{1,2}\|'_{Q_{1,2},a,b}<\infty$ for $a\geq
a_1$, $b>b_1$ sufficiently large, i.e., $f_{1,2}\in
S^0_\beta(Q_{1,2})$.
The following lemma allows to apply the H\"ormander's
$L_2$-estimates~\cite{H}
to prove of the existence of a function
$\psi$ with the specified properties.

\lemma
\label{l8}
{\it For any $a,b>0$ and any nonempty cone $U\subset
\R^k$ there are a plurisubharmonic function $\rho$ on
$\C^k$, positive numbers $a',b'$, and a constant $H$ such that $\rho_{U,a,b}-H\leq \rho\leq
\rho_{U,a',b'}$.}

Supposing Lemma~\ref{l8} is proved, we finish the derivation of
Lemma~\ref{l7}. Let $U=Q_1\cup Q_2$.  By~(\ref{a0}) and the condition of
Lemma~\ref{l7}, there are $a,b>0$ such that $\|\eta_j\|'_{U,a,b}<\infty$.
According to Theorem~4.4.2 of~\cite{H} there exists a solution $\psi$
of equations~(\ref{a1}) such that
\begin{equation}
2\int_{\C^k}|\psi|^2
e^{-\rho}(1+\|w\|^2)^{-2}\,\di\lambda(w)\leq \sum_{j=1}^k\int_{\C^k}
|\eta_j|^2 e^{-\rho}\di\lambda(w),
\label{a2}
\end{equation}
where
$\|w\|=(|w_1|^2+\ldots+|w_k|^2)^{1/2}$.
By Lemma~\ref{l8},
the integrals in the right-hand
side are bounded
by $e^H(\|\eta_j\|'_{U,a,b})^2$
and, therefore, are convergent.
Estimating
$e^{-\rho}$ in the left-hand side of (\ref{a2}) from below by
the function
$e^{-\rho_{U,a',b'}}$, we conclude that $\|\tilde\psi\|'_{U,a',b'}<\infty$,
where $\tilde\psi=(1+\|w\|^2)^{-1}\psi$. To complete the proof, it remains
to note that $\|\psi\|'_{U,a_1,b_1}\leq
C\|\tilde\psi\|'_{U,a',b'}$ for $a_1> a'$, $b_1>b'$.

{\it Proof of Lemma~$\ref{l8}$}
is essentially contained in the derivation of Theorem~5
of~\cite{Sol2}. We assume $0<a<1/2ek$; to pass to the general case,
it suffices to make a rescaling of the arguments. Let $\sigma=eka$.
By Lemma~4 of~\cite{Sol2}, there are a sequence
$\varphi_N(w)\in S^0_\beta(\R)$ and constants
$A,B>0$ independent of $N$ such that
\begin{align} & |\varphi_N(w)|\leq
A\exp(|q|-|p/b|^{1/\beta}),\quad w=p+iq\in \C, \label{a3}\\
& \ln
|\varphi_N(iq)|\geq \sigma |q|,\label{a4}\\ & \ln |\varphi_N(w)|\leq
|q|-N\ln^+(\sigma|p|/N)+B,
\label{a5}
\end{align}
where $\ln^+ r=\max(0,\ln
r)$. Let
$a'\geq 2$ and let
\begin{equation}
\tilde\rho(w) = \sup_{\kappa\in
\R^k,N}\{\Phi_N(w-\kappa)+M_N(\kappa)\}, \,\,\, M_N(\kappa) =
\inf_{w\in\C^k}\{-\Phi_N(w-\kappa)+\rho_{U,a',b}(w)\},
\label{a6}
\end{equation}
where $\Phi_N(w)=2\sum_{j=1}^k \ln |\varphi_N(w_j)|$.
Obviously, we have $\tilde \rho\leq \rho_{U,a',b}$. Since functions $\Phi_N$
are plurisubharmonic,
$\rho(w)=\varlimsup_{w'\to w}\tilde\rho(w)$
is also a plurisubharmonic function, see~\cite{V}, Sec.~II.10.3. In view of
the continuity of
$\rho_{U,a',b}$ we have $\tilde\rho\leq\rho\leq\rho_{U,a',b}$
and it remains to show that $\tilde\rho\geq\rho_{U,a,b}-H$.
From (\ref{a3}) and the inequality $|p_j-\kappa_j|^{1/\beta}\geq
|p_j|^{1/\beta}-|\kappa_j|^{1/\beta}$,
it follows that
\begin{equation}
-\Phi_N(w-\kappa)/2-\sum\nolimits_j |p_j/b|^{1/\beta}+ \sum\nolimits_j|q_j|
\geq -k\ln A -\sum\nolimits_j |\kappa_j/b|^{1/\beta},
\nonumber
\end{equation}
and hence
$M_N(\kappa)\geq -k\ln A -\sum\nolimits_j
|\kappa_j/b|^{1/\beta}+L_N(\kappa)$, where
\begin{equation}
L_N(\kappa)=\inf_{w\in \C^k}\left\{
-\Phi_N(w-\kappa)/2+\inf_{p'\in
U}\sum\nolimits_j |p_j-p'_j|+\sum\nolimits_j|q_j|\right\}.
\nonumber
\end{equation}
Therefore,
estimating
the supremum in $(\ref{a6})$ from below by the value of the function at
$\kappa=p=\mathrm{Re}\,w$
and taking~(\ref{a4}) and the inequality
$2\sigma>a$ into account,
we find that
$$ \tilde\rho(w)\geq a\sum\nolimits_j
|q_j|-\sum\nolimits_j |p_j/b|^{1/\beta}+\sup_N L_N(p)-k\ln A.  $$
Thus, it
suffices to show that $\sup_N L_N(p)\geq a\inf_{p'\in U}\sum\nolimits_j
|p_j-p'_j|-C$. Passing to the Euclidean norm $\|p\|$, using the
elementary inequalities $\sum_j \ln^+ |p_j|\geq \ln^+ (\|p\|/\sqrt{k})$ and $\|p\|\leq
\sum_j |p_j|$, and estimating $\Phi_N$ by~(\ref{a5}), we conclude
that
\begin{equation}
L_N(p) \geq -kB + \inf_{\lambda\in \R^k} \{N\ln^+
(\sigma \|\lambda\|/N\sqrt{k})
+\delta_U(p+\lambda)\},
\nonumber
\end{equation}
where $\delta_U(p)=\inf_{p'\in U}\|p-p'\|$.
Estimating $\delta_U(p+\lambda)$ from below by
$\max(\delta_U(p)-\|\lambda\|,0)$ and calculating the infimum
with respect to $\lambda$
yield $L_N(p)\geq N\ln^+(\sigma \delta_U(p)/N\sqrt{k})-kB$.
Let $\delta_U(p)\geq e\sqrt{k}/\sigma$ and let $N_0$ be
the integer part of $\sigma\delta_U(p)/e\sqrt{k}$. In view of
the inequality $\sum_j |p_j|\leq \sqrt{k}\|p\|$ we find that
$$
\sup_N
L_N(p)\geq L_{N_0}\geq \frac{\sigma\delta_U(p)}{e\sqrt{k}}-1-kB\geq a
\inf_{p'\in U} \sum_j|p_j-p'_j| -C.
$$
Since $L_N(p)\geq -k\ln A$ by (\ref{a3}),
this inequality holds for all $p\in \R^k$ with a new
constant $C$.
The lemma is proved.

\section*{Appendix~B. Proof of Lemma~\ref{l4}}

By Lemma~2 of~\cite{Komatsu}, the spaces $L^{(1,2)}$
are representable as inductive limits of sequences of Banach spaces
$L_k^{(1,2)}$ with injective connecting mappings
$u^{(1,2)}_{km}\,:\,L^{(1,2)}_k\to L^{(1,2)}_m$, $1\leq k\leq m$,
which take unit balls in
$L^{(1,2)}_k$ to compact subsets of $L^{(1,2)}_m$. Let
$M_k=L^{(1)}_k\otimes_i L^{(2)}_k=L^{(1)}_k\otimes_\pi L^{(2)}_k$ and
$M=\varinjlim_{k} M_k$. We denote by $u_k^{(1,2)}$ and $u_k$ the canonical
embeddings of $L^{(1,2)}_k$ into $L^{(1,2)}$ and of $M_k$ into $M$
respectively.
One can identify the space $M$ with $L^{(1)}\otimes_i
L^{(2)}$
using
the canonical separately continuous bilinear mapping from
$L^{(1)}\times L^{(2)}$ into $M$ which is uniquely determined by
the relations
\begin{equation}
u^{(1)}_k(x_1)\otimes u^{(2)}_k(x_2) = u_k(x_1\otimes x_2),\quad x_{1,2}
\in L_k^{(1,2)}.
\label{5}
\end{equation}
To prove the lemma, it suffices to show that this mapping is continuous.
Let $V$ be an absolutely convex neighborhood of the origin in $M$.
Set $V_k=u_k^{-1}(V)$.
We shall construct sequences of absolutely convex
neighborhoods $V_k^{(1,2)}$ of the origin in $L^{(1,2)}_k$ such that
\begin{enumerate}
\item[(i)] $u^{(1,2)}_{km}(V_k^{(1,2)})\subset V_m^{(1,2)}$ for $m>k$;
\item[(ii)] $V_k^{(1)}\otimes V_k^{(2)} \subset V_k$, $k=1,2\ldots$;
\item[(iii)] The set $u^{(1,2)}_{km}(V_k^{(1,2)})$ is compact in
$L^{(1,2)}_m$ for $m>k$.
\end{enumerate}
The sets $V^{(1,2)}=\bigcup_{k=1}^{\infty} u^{(1,2)}_k (V^{(1,2)}_k)$
are neighborhoods of the origin in $L^{(1,2)}$ because they are absolutely convex
(in view of (i)) and $[u^{(1,2)}_k]^{-1}(V^{(1,2)})$ contain $V^{(1,2)}_k$.
Moreover, by (\ref{5}) and property (ii), we have $V^{(1)}\otimes
V^{(2)}\subset V$, i.e., the mapping $(x,y)\to x\otimes y$ from
$L^{(1)}\times L^{(2)}$ to $M$ is continuous.

We construct the sequences $V_k^{(1,2)}$ by induction.
Let $\tilde V_k$ denote the inverse image of $V_k$ under the canonical
bilinear mapping from $L^{(1)}_k\times L^{(2)}_k$ to
$M_k$. Let $V_1^{(1,2)}$ be closed balls in $L^{(1,2)}_1$ such that
$V_1^{(1)}\times V_1^{(2)}\subset \tilde V_1$. Suppose
$V^{(1,2)}_1,\ldots,V^{(1,2)}_k$ satisfying (i)--(iii) are constructed.
The compactum $u^{(1)}_{k,k+1}(V^{(1)}_k)\times u^{(2)}_{k,k+1}(V^{(2)}_k)$
is contained in the open neighborhood of the origin $\tilde V_{k+1}$.
Hence, there are closed balls $B^{(1,2)}$ in
$L^{(1,2)}_{k+1}$ such that $ [u^{(1)}_{k,k+1}(V^{(1)}_k)+B^{(1)}]\times
[u^{(2)}_{k,k+1}(V^{(2)}_k)+B^{(2)}]\subset \tilde V_{k+1}$.
Set $V^{(1,2)}_{k+1}=u^{(1,2)}_{k,k+1}(V^{(1,2)}_k)+B^{(1,2)}$.
Conditions (i) and (ii) are obviously satisfied. If $m> k+1$, then
$u^{(1,2)}_{k+1,m}(V^{(1,2)}_{k+1})$ is the sum of the compact sets
$u^{(1,2)}_{k,m}(V^{(1,2)}_{k})$ and $u^{(1,2)}_{k+1,m}(B^{(1,2)})$ and,
therefore, is also compact.

To prove the second statement of the lemma it suffices to note that
$i$- and $\pi$-topologies coincide on the tensor product of the
Fr\'echet spaces
$L^{(1)\prime}$ and $L^{(2)\prime}$ and that
$(L^{(1)}\hat\otimes_\pi L^{(2)})'=L^{(1)\prime}\hat\otimes_\pi
L^{(2)\prime}$ for arbitrary DF-spaces $L^{(1,2)}$ one of which is
nuclear (see~\cite{Schaefer}, Chapter IV, Problem~32). The lemma is proved.

\end{document}